%% file: cg_trail.tex
\documentclass[doublespacing]{elsartx}

\usepackage{icarus}

\usepackage{natbib}

\bibpunct{(}{)}{;}{a}{,}{,}

\usepackage[]{graphicx}

\usepackage[nolists,tablesfirst]{endfloat}
\usepackage{array}
\usepackage{tabularx}
\usepackage{lscape} 
\usepackage{color}
\usepackage{colortbl}

\newcommand{\BibTeX}{ \textrm{B\kern-.05em\textsc{i\kern-.025em b}\kern-.08em
    T\kern-.1667em\lower.7ex\hbox{E}\kern-.125emX} }
\newcommand{\arcmin}{$^{\prime}$}
\newcommand{\arcsec}{$^{\prime\prime}$}
\newcommand{\degree}{$^{\circ}$}
\newcommand{\micron}{$\mu{\textrm m}$}

\newcommand{\df}{{\rm d}}

\newcommand{\te}{t_{\rm e}}
\newcommand{\tobs}{t_{\rm {obs}}}
\newcommand{\ve}{v_{\rm e}}
\newcommand{\Psij}{\mathbf \Psi_{\mathbf j}}
\newcommand{\tilPsij}{\tilde{\mathbf \Psi}_{\mathbf j}}
\newcommand{\Tj}{\mathbf T_{\mathbf j}}
\newcommand{\afr}{$Af\!\rho$}
\newcommand{\cg}{CG}

\newcommand{\vr}{\mathbf{r}}
\newcommand{\vrref}{\mathbf{r}^{(i)}_{\mathrm {ref}}}
\newcommand{\vve}{\mathbf{v_{\mathrm {e}}}}
\newcommand{\vvref}{\mathbf{v}^{(i)}_{\mathrm {ref}}}
\newcommand{\Phii}{{\mathbf \Phi}^{(i)}}

\newcommand{\Isun} {I_{\odot}}

\newcommand{\Msun} {M_{\odot}}

\newcommand{\gr}{\cellcolor[gray]{.7}}

\begin{document}

\begin{frontmatter}



\title{The Dust Trail of Comet 67P/Churyumov-Gerasimenko between 2004 and 2006}


\author[agarwal]{Jessica Agarwal}, 
\author[mueller]{Michael M{\"u}ller},
\author[reach]{William T. Reach}, 
\author[sykes]{Mark V. Sykes},
\author[boehnhardt]{Hermann Boehnhardt}, and
\author[gruen1,gruen2]{Eberhard Gr{\"u}n}

\address[agarwal] {European Space Research and Technology Centre, 
                                2200 AG Noordwijk (The Netherlands)}
\address[mueller] {European Space Operations Centre,
                                64293 Darmstadt (Germany)}
\address[reach] {Infrared Processing and Analysis Center, MS 220-6, California Institute of
  Technology, Pasadena, CA 91125 (U.S.A.)}
\address[sykes]{Planetary Science Institute,
				Tucson, AZ 85719 (U.S.A.)}
\address[boehnhardt]{Max Planck Institute for Solar System Research, 
				37191 Katlenburg-Lindau (Germany)}
\address[gruen1]{Max Planck Institute for Nuclear Physics,
				69117 Heidelberg (Germany)}
\address[gruen2]{Laboratory for Atmospheric and Space Physics 
University of Colorado, Boulder, CO, 80303-7814 (U.S.A.)}



%
%
%
%
%


\end{frontmatter}



\begin{flushleft}
\vspace{1cm}
Number of pages: \pageref{lastpage} \\
Number of tables: 6\\
Number of figures: 13\\
\end{flushleft}


\begin{pagetwo}{The Dust Trail of 67P/Churyumov-Gerasimenko 2004 - 2006}

Jessica Agarwal \\
ESA/ESTEC\\
Keplerlaan 1\\
2201 AZ Noordwijk\\
The Netherlands\\
\\
Email: jagarwal@rssd.esa.int\\
Phone: +31 71 565 4928 \\
Fax: + 31 71 565 4697
\end{pagetwo}

\begin{abstract}
We report on observations of the dust trail of comet 67P/Chur\-yu\-mov-Gera\-si\-menko (\cg) in visible light with the Wide Field Imager at the ESO/MPG 2.2m telescope at 4.7\,AU before aphelion, and at 24\,\micron\ with the MIPS instrument on board the Spitzer Space Telescope at 5.7\,AU both before and after aphelion. The comet did not appear to be active during our observations. Our images probe large dust grains emitted from the comet that have a radiation pressure parameter $\beta < 0.01$. 
We compare our observations with simulated images generated with a dynamical model of the cometary dust and constrain the emission speeds, size distribution, production rate and geometric albedo of the dust. We achieve the best fit to our data with a differential size distribution exponent of -4.1, and emission speeds for a $\beta = 0.01$ particle of 25 m/s at perihelion and 2 m/s at 3 AU. The dust production rate in our model is on the order of 1000 kg/s at perihelion and 1 kg/s at 3 AU, and we require a dust geometric albedo between 0.022 and 0.044.
The production rates of large ($>$10\,\micron) particles required to reproduce the brightness of the trail are sufficient to also account for the coma brightness observed while the comet was inside 3\,AU, and we infer that the cross-section in the coma of \cg\ may be dominated by grains of the order of 60-600\,\micron.
\end{abstract}



\section{Introduction}
\label{sec:intro}

A cometary trail is a narrow and elongated structure along the projected orbit
of a comet. Trails are thought to consist of dust particles characterised by
little sensitivity to solar radiation pressure and emitted at low speeds
relative to the comet, such that they remain close to the orbit of the parent comet for
many revolutions around the Sun. 
Such properties are commonly associated with mm- to
cm-sized particles, and the trail is therefore interpreted as a 
continuous record of the emission history of such particles from the comet
(\cite{sykes-walker1992a}). 
The tail of a comet, by contrast, consists of smaller dust particles, which
are more sensitive to radiation pressure and are therefore removed from the
cometary environment on timescales of weeks to months. The tail roughly points
away from the nucleus in the direction opposite to the Sun, while the trail is
found along the orbit of the comet. 
Since the transition from ``small'' to ``large'' particles is likely a
smooth one, the distinction between tail and trail is not always clear,
in particular when the observer is close to the orbital plane of the comet. In
such a configuration, also a neckline can become apparent, which is an accumulation
of dust emitted 180\degree\ in true anomaly before the observation
(\cite{kimura-liu1977}). 

During most available observations of the 67P/Churyumov-Gerasimenko trail
(with the exception of 
those described by \cite{ishiguro2008}), the viewing geometry did not favour a clear
distinction between the trail on the projected orbit and smaller tail particles
 that in
3D were outside the orbit, but appeared in projection almost under the same
position angle as the orbit. It is therefore difficult to discriminate between
(smaller) dust emitted during the previous perihelion passage and (larger) dust emitted at least
one orbital period before the observation. 
From our modelling we expect that most of the
dust seen in our images close to the projected orbit was emitted
during the last perihelion passage. 
Nevertheless, we will in this paper use the expression ``trail'' for all dust
observed close to the projected orbit of the comet, while employing more
specific terms if referring to dust emitted during a more restricted time
interval. 
 

The dust trail of comet CG has been more closely
studied than most other comet trails, mainly because CG is the target of the Rosetta
mission of the European Space Agency (\cite{glassmeier-boehnhardt2007}). 
The trail was discovered with the Infrared Astronomical Satellite (IRAS) in 1983 
at 12, 25, and 60\micron, and over a length ranging from 0.1\degree\ in mean anomaly
ahead of the nucleus to 1.1\degree\ behind it (\cite{sykes-hunten1986,sykes-lebofsky1986,sykes-walker1992a}). 
The next observations of the CG trail were made between
September 2002 and February 2003 with the KISO telescope at visible
wavelengths (\cite{ishiguro2008}). Since this was immediately after
perihelion in August 2002, the comet was active and displayed a tail
of young particles. The
position angles of the projected tail and trail differed significantly, such
that the old particles along the orbit were well separated from younger and
smaller particles.
In March 2003, the position angles of the (faint) trail of old
particles and of smaller particles emitted during the 2002 apparition differed
by a few degrees, and the old trail was not
detected in a visible image made with the Telescopio Nazionale Galileo on 27
March 2003 (\cite{fulle-barbieri2004a,moreno-lara2004}). \cite{fulle-barbieri2004a} interpret the observed extended dust
structure in terms of a neckline and infer that the comet emitted a
significant amount of large particles already at 3.5~AU before
perihelion, which has been controversial ever since
(\cite{agarwal-mueller2007}). 
\cite{kelley-reach2008} have observed the CG trail at 3.2\,AU in
June 2003 (optical, from Mount Palomar) and at 4.5\,AU in February 2004 (24\,\micron\ with
Spitzer). While the Palomar image was dominated by the
neckline-structure, the older trail could be separated
from the neckline in the Spitzer data.
The nucleus of \cg\ was closely monitored with ESO/VLT/FORS2 and the three instruments of Spitzer around aphelion between 2004 and 2007 by \cite{tubianaPhD, tubiana-barrera2008, kelley-wooden2009}. Some of these observations were sufficiently sensitive to detect also the trail within a few arcminutes from the nucleus. In visible light the surface brightness of the trail was measured in April 2004, June 2004, and May 2006 at heliocentric distances of 4.7, 4.9, and 5.6\,AU (\cite{tubianaPhD, tubiana-barrera2008}). The June 2004 observation was performed in R, V, and I filter, resulting in the first-ever determination of the visible colours of the trail. In the infrared, the trail was observed at 24\,\micron\ in August/September 2006 (5.5\,AU out-bound) and in May 2007 (4.8\,AU in-bound) by \cite{kelley-wooden2009}. Observations of the \cg\ trail are summarised in Table~\ref{tab:observations_summary}.

\begin{table}[h]
\caption{Observations of the \cg\ trail and/or neckline. The heliocentric distance $r_{\rm h}$ is given in AU. ``Kiso'' refers to the Kiso Observatory 1.05m Schmidt telescope at Nagano, Japan; ``TNG'' is the Telescopio Nazionale Galileo at the Observatorio Roque de los Muchachos on La Palma, Spain; ``Palomar'' refers to the Mount Palomar 5m Hale telescope Large Format Camera; ``MIPS'' is the Multiband Imaging Photometer for Spitzer; ``FORS2'' is the FOcal Reducer and low dispersion Spectrograph at ESO/VLT; and ``WFI'' is the Wide Field Imager at the ESO/MPG 2.2m telescope on La Silla. 
$\Delta$MA is the range in mean anomaly corresponding to the fields of view of the observations, except for IRAS, where the range of the detected trail is given.}
%
\centering
\label{tab:observations_summary}
\begin{tabular}{llllll}
\hline\noalign{\smallskip}
Date & $r_{\rm h}$ & Instrument &Wavelength & $\Delta$MA [\degree] & Reference\\
\hline\hline\noalign{\smallskip}
May 1983 & 2.3 & IRAS & 12,25,60 $\mu$m & 0.1 .. -1.1 & \cite{sykes-lebofsky1986,sykes-hunten1986},\\
&&&&&\cite{sykes-walker1992a}\\
Sep 2002 & 1.3 & Kiso  & R & 0.0 .. -0.2 & \cite{ishiguro2008}\\
Dec 2002 & 1.8    & Kiso   & R & 0.1 .. -0.3 & \cite{ishiguro2008}\\
Feb 2003 & 2.2 & Kiso   & R & 0.1 .. -0.2 & \cite{ishiguro2008}\\
Mar 2003 & 2.6 & TNG   & R &0.0 .. -0.1& \cite{fulle-barbieri2004a},\\
&&&&&\cite{moreno-lara2004}\\
Jun 2003 & 3.2 & Palomar   & R     & 0.3 .. -0.6 & \cite{kelley-reach2008}\\
Feb 2004 & 4.5 & MIPS   & 24 $\mu$m   & 0.1 .. -0.4 & \cite{kelley-reach2008}\\
Apr 2004 & 4.7 & FORS2   & R  & 0.0 .. -0.1& \cite{tubianaPhD},\\
&&&&&\cite{tubiana-barrera2008}\\
Apr 2004 & 4.7 & WFI & vis, no filter & 0.1 .. -1.0 & \cite{agarwal-boehnhardt2007},   \\
&&&&&this paper\\
Jun 2004 & 4.9 & FORS2   & R,V,I     & 0.1 .. -0.1 & \cite{tubianaPhD},\\
&&&&&\cite{tubiana-barrera2008}\\
Aug 2005 & 5.7 & MIPS   & 24 $\mu$m       & 1.0 .. -1.6 & this paper\\
Apr 2006 & 5.7 & MIPS   & 24 $\mu$m       & 0.5 .. -1.7 & this paper \\
May 2006 & 5.6 & FORS2   & R   & 0.1 .. -0.1 & \cite{tubianaPhD},\\
&&&&&\cite{tubiana-barrera2008}\\
Sep 2006 & 5.5 & MIPS   & 24 $\mu$m       & 0.2 .. -0.2 & \cite{kelley-wooden2009} \\
May 2007 & 4.8 & MIPS   & 24 $\mu$m        & 0.1 .. -0.1& \cite{kelley-wooden2009}\\

%
\\
\noalign{\smallskip}\hline
\end{tabular}
\end{table}

In this paper we discuss data from three observations of the CG trail taken
at heliocentric distances beyond 4.7 AU. The
first observation was carried out in visible light with the Wide Field Imager
(WFI) at the ESO/MPG 2.2m telescope in April 2004. It was followed in August
2005 and April 2006 by two observations at 24\micron\ with Spitzer/MIPS. 
The comet did not appear to be active during our observations, which is consistent with \cite{tubiana-barrera2008} reporting a point-source like nucleus during all their observations between June 2004 and August 2006. Due to the large
heliocentric and geocentric distances, the employed fields of view (FOVs) of about half a degree
in size allowed us to cover a comparatively large part of the trail (up to two
degrees in mean anomaly). The data acquisition and processing are described in
Section~\ref{sec:obs} of this paper. 
The purpose of acquiring these data was to obtain a solid observational basis
to model the past emission of mm- to cm-sized particles from CG, extending the
previously existing data series both in temporal and spatial coverage. It is
problematic to infer the production rate of such large particles directly from
observations of the coma when the comet is active, because in that situation large and small particles are mixed.

Our aim was to find a model (in terms of production rate, size
distribution, and emission speed of large
particles) that would enable us to reproduce the observed time-evolution of
the trail. 
To simulate images of the dust environment for a given set of parameters, we
use a generalisation of the method introduced by \cite{finson-probstein1968a} such that the results are valid for
the longer timescales required to simulate a trail. The method is outlined at
the beginning of
Section~\ref{sec:model}, and the mathematical details can be found in
Appendix~\ref{app:math}. 
In Section~\ref{sec:model} we also present the physical model employed for the
cometary dust production, and introduce the variable parameters we aim to constrain. 
Simulated images for a range of parameters are shown in
Section~\ref{sec:results}. We discuss the impact of each parameter on the
morphology of the images. By comparison to the observations we derive the
best-fitting set of parameters.
We discuss our results in Section~\ref{sec:discussion}.

\section{Observations and Data Reduction}
\label{sec:obs}
We have observed the dust trail of comet \cg\ at three points on the outer
section of its orbit, once in visible light and twice at 24\,\micron.
Parameters of these observations are listed in
Table~\ref{tab:obs_geom}. In this section, we describe the 
processing applied to each of the data sets, and derive calibrated profiles
of the surface brightness, and the trail width along the orbit.
\begin{table}[h]
\caption{Geometrical parameters of the three observations of the dust trail
of comet \cg.
MA = mean anomaly, PA = position angle.
%
The ``orbit plane angle'' is the angle between observer and comet orbital
plane, measured from the centre of the comet nucleus; positive values indicate
that the observer is
above the comet orbital plane, in the direction of the orbital angular momentum
vector of the comet. 
The ``viewing angle'' is the angle between the line of sight and
the comet velocity vector; values $<$ 90\degree\ indicate that the comet moves
away from the observer. 
The ``neckline emission date'' is defined as the date when the comet nucleus
was at a true anomaly of $\tau_{\rm obs} \,-$\,180\degree. 
With the exception of the orbit section lengths and the viewing angle, all
values were obtained from the JPL Horizons System ({\tt
http://ssd.jpl.nasa.gov/horizons.cgi}).} 
\begin{center}
\label{tab:obs_geom}  
\begin{tabular*}{\textwidth}
{@{\extracolsep{\fill}}lrrr}
\hline\noalign{\smallskip}
Telescope                                 &ESO/MPG 2.2m                   &Spitzer                   &Spitzer\\
Instrument                                &WFI                            &MIPS24                    &MIPS24\\
Date of observation$^\dagger$       &18--21 Apr 2004         & 28/29 Aug 2005              &8/9 Apr 2006\\ 
\hline\hline\noalign{\smallskip}
Heliocentric distance                     &       4.69 AU                 &      5.69 AU             &    5.66 AU\\
Distance from observer                    &       3.69 AU                 &      5.72 AU             &    5.36 AU\\
Orbit section covered (proj.)             & 2\arcmin \ldots $-$33\arcmin    & 18\arcmin \ldots $-$28\arcmin  & 9\arcmin \ldots $-$33\arcmin\\
Orbit section covered (MA)                & 0.1\degree \ldots $-$1.0\degree & 1.0\degree \ldots $-$1.6\degree& 0.5\degree \ldots $-$1.7\degree\\
PA of neg. velocity vector                & 296.9\degree                  & 293.5\degree             & 281.6\degree\\
PA of Sun-comet vector                    & 253.1\degree                  & 106.2\degree             &274.9\degree\\
Phase angle                               & 1.0\degree                    & 10.3\degree              &10.0\degree\\
Orbit plane angle                         & \,$-$0.7\degree                 &1.3\degree                &\,$-$1.2\degree\\
Viewing angle                             & 54\degree                     &94\degree                 &110\degree\\
True anomaly at obs. ($\tau_{\rm obs}$)    & 150.8\degree                  &176.1\degree              &185.9\degree\\
Neckline emission date ($t_{\rm nl}$ )  & 15 Jul 2002                   &14 Aug 2002               &25 Aug 2002\\
$t_{\rm nl}$ relative to perihelion& $-$34 d                         & $-$4 d                     &+7 d\\
Heliocentric distance at $t_{\rm nl}$      &1.36 AU                        &1.29 AU                   &1.29 AU\\
\noalign{\smallskip}\hline
\end{tabular*}
\end{center}
\begin{small}
$^\dagger$The values given in this table are for 20 Apr 2004, 29 Aug 2005, and 9 Apr 2006, 00:00:00.0 UT.
\end{small}
\end{table}

\subsection{Wide Field Imager Observation}
\label{subsec:obs_wfi}
Comet \cg\ was observed in April 2004 with
the Wide Field Imager (at the ESO/MPG 2.2m telescope on La Silla
(Chile). The 
heliocentric and geocentric distances of the comet were \mbox{4.7 AU} and
\mbox{3.7 AU}, respectively.  
The camera has eight Charge Coupled Devices (CCDs).
Each image is therefore a mosaic of eight frames covering a total FOV of
\mbox{34\arcmin$\times$33\arcmin}. 
The data set includes 60 exposures of \mbox{540\,s} exposure
time each, of which 15 were made during each night between the 18 and 21 April 2004. We did not use data from the easternmost four CCDs from the first night because they were contaminated by straylight from a magnitude 4 star.
The total integration time was \mbox{9\,h} in the western part of the image, and \mbox{6.8\,h} in the eastern part.
To maximise sensitivity, the observations were made without filter 
and using 3$\times$3 on-chip pixel binning, resulting in a pixel size of 0.71\arcsec.

The details of the data processing are described in
\cite{agarwal-boehnhardt2007} and \cite{agarwalPhD} and are briefly summarised as
follows. After bias-subtraction, all eight exposures obtained simultaneously
by the different CCDs were combined into a single mosaic, and further processing was done on these mosaics. Superflats were derived from the science
data by normalising and median-combining five consecutive exposures
each. Division by the superflat corrected also gain differences between
individual CCDs and fringing effects. The flatfielded images were
airmass-corrected using a mean extinction coefficient for La Silla of
0.15\,mag/airmass, and the mean sky level (approximated by the mode of the
image) was subtracted.
Object masks were obtained by average-combining all images of one night in the
rest frame of the stars and masking all objects brighter than five times the
local mean variance of the sky, using the IRAF  routine {\tt objmasks}. 
The
final image of the trail was obtained by average-combining all images in the
rest frame of the comet while applying the object masks and interpolating the individual exposures to a common frame using the Spitzer Science Center's MOPEX software (\cite{makovoz-khan2005}). In the
resulting image the trail is visible, but the low average signal-to-noise
ratio precludes quantitative analysis. The image was therefore smoothed 
replacing each pixel by the average over a rectangular neighbourhood of
140\arcsec\ parallel and 7\arcsec\ perpendicular to the trail.
\begin{figure}[t]
%
\includegraphics[clip,width=\textwidth]{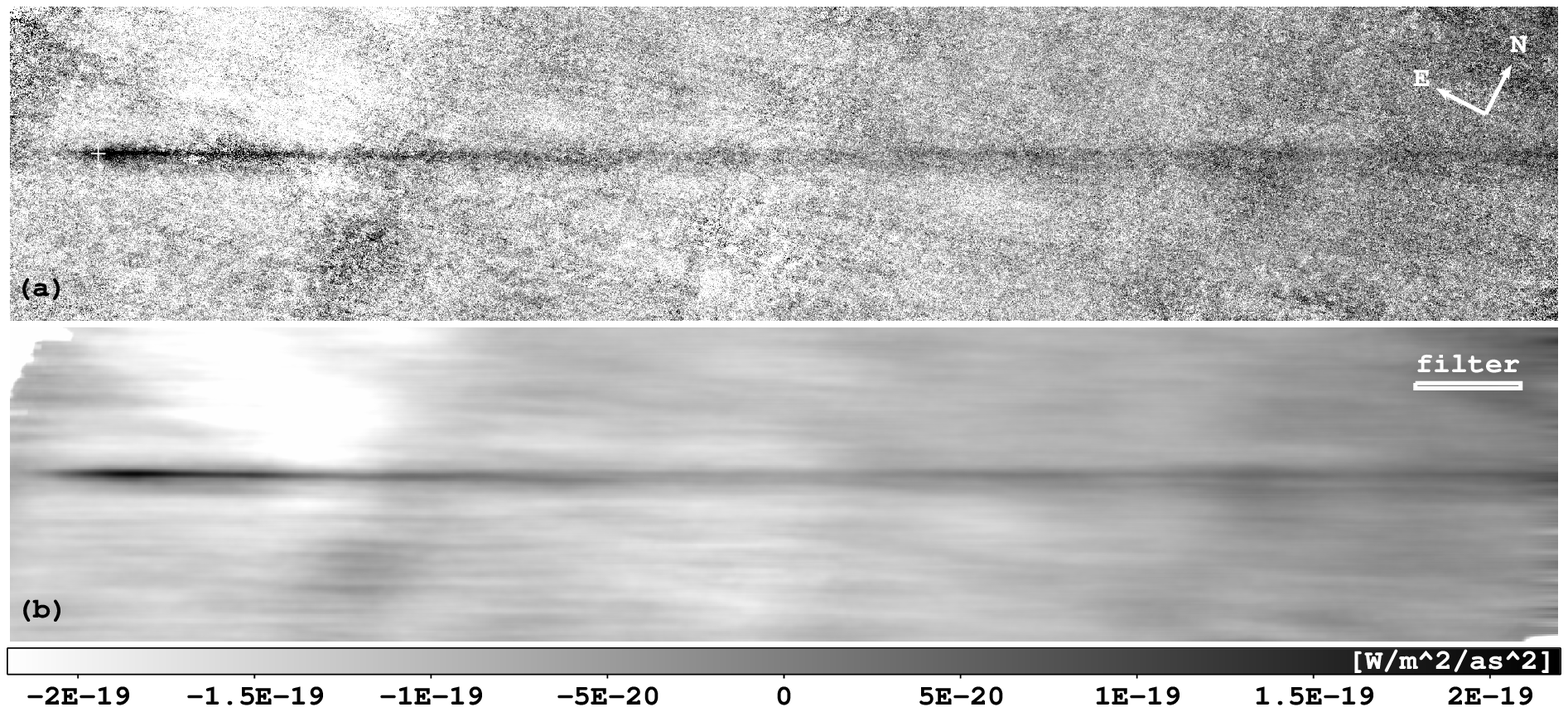}
\includegraphics[clip,width=\textwidth]{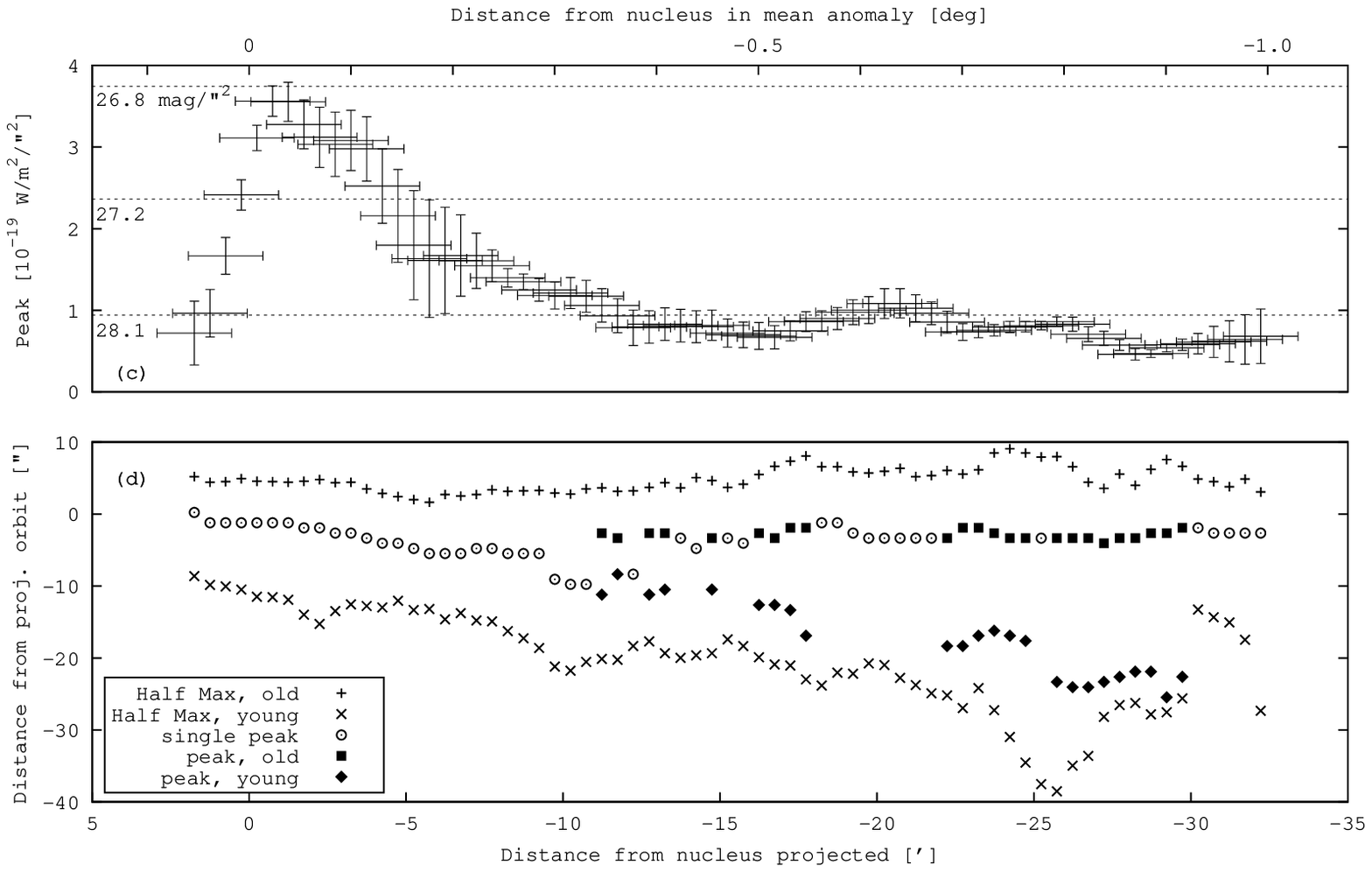}
\caption{Optical image of comet \cg\ obtained in April 2004 with the Wide Field Imager at the ESO/MPG 2.2m telescope on La Silla. 
  The four panels are aligned in x, the scale in projected distance from the nucleus is given at the bottom of panel d, and the corresponding offset in mean anomaly of the comet is shown at the top of panel c.
One arcminute corresponds to 1.6$\times$10$^8$\,m at the comet.
  {\bf a)} Unfiltered mosaic image.
  {\bf b)}~Same image, after replacing each pixel by the average over a
  neighbourhood of 200 pixels (140\arcsec)
  parallel and 10 pixels (7\arcsec) perpendicular to the trail
  axis (boxcar-filtering). The size of the filtering window is
  indicated in the upper right corner. 
  The data points  in panels (c) and (d) were derived from profiles perpendicular to the projected orbit at discrete distances $x$ from the nucleus. Examples of such profiles are displayed in panels (c) to (f) in Fig.~\ref{fig:best_fit_04}.
  {\bf c)} Height above background of the brightest point in the trail cross-section. The x-errorbars are the size of the filter window. The y-errorbars are the rms of the background.
  Dashed lines indicate the corresponding surface brightness in R-filter. 
  {\bf d)} ``+'' and ``x'' show
  the distances from the orbit where the surface brightness is half of the
  peak
  brightness,``+'' indicating the trail flank that is closer to Sun and observer.  
  Circles show the peak position 
  when only one peak is observed. Otherwise, squares and diamonds indicate the
  positions of the peaks attributed to old trail and neckline, respectively (see beginning of Sec.~\ref{sec:results}).
}
\vspace{-2\baselineskip}
\label{fig:wfi_images}
\end{figure}

Flux calibration was achieved through aperture photometry of solar type field 
stars
within the field of view of the images. Stars were considered as ``solar type''
when their B-R and R-I filter colours in the USNO-B1.0 catalogue
\citep{monet-levine2003} were compatible with solar values. 
To achieve a comparison with simulated images, we assumed a solar R-magnitude
of -27.1 corresponding to a flux of 1367 W m$^{-2}$ (\cite{cox2000}). 

The statistical uncertainty of this calibration is 0.02 mag, which is an order of magnitude
smaller than the uncertainty of a single entry in the catalogue, due to the
large sample of 200 data points employed for the calibration. This
uncertainty, however, is negligible compared to the one introduced by the
variation in the background, which is responsible for a relative uncertainty
in the measured flux of up to 10\% in the faint part of the trail (see
errorbars in Fig.~\ref{fig:wfi_images}c). 
A systematic error was introduced by the process of superflatfielding.
Trail information could not be completely excluded from the superflats due to
the employed jittering pattern and because the trail signal was fainter than
the statistical variation of the background in the raw data. 
To nonetheless compare the observation with
model images, the flatfielding and smoothing process was included in the
simulation.

\cite{tubianaPhD} has observed the \cg\ trail within several arcminutes from the nucleus on 14 April 2004 with the FORS2 instrument at ESO/VLT using R-filter. Cross-cut profiles through the trail at 24.25\arcsec\ behind the nucleus from their VLT and our WFI observations are in good agreement around the projected orbit and north of it, which confirms the accuracy of the calibrations, done independently from each other. The WFI image shows additional brightness south of the projected orbit, which may be due to background- or fringing-contamination.

The trail in our April 2004 image is characterised by a bright bulge stretching over 5\arcmin\ behind the nucleus. Beyond this region, the
brightness is approximately constant up to the edge of the field of view
nearly 35\arcmin\ behind the nucleus. The width of the trail increases
steadily with distance from the nucleus, and at distances larger than
11\arcmin\ a double peaked structure is observed
(Fig.~\ref{fig:wfi_images}d). 
Since brightness variations of similar scales are present in the background, we cannot exclude that the double peak is an artefact. The difference
in position angle of the two peaks is 0.8\degree $\pm$ 0.2\degree, which agrees with the expected position angles of neckline and older trail (see Section~\ref{sec:results}).

To assess whether we can detect on-going or recent cometary activity, we have compared cross-cut profiles through the nucleus position with similar profiles through asteroids in the FOV (\cite{boehnhardt-delahodde2002}). The accuracy of this method is in our case low, because the pixel scale of 0.71\arcsec\ just barely resolves the seeing disc that varied between 0.5\arcsec\ and 2\arcsec\ during the observations. Within this limitation, we did not find any deviation of the nucleus profile from that of a point source and thus did not detect cometary activity.

\subsection{Spitzer/MIPS Observations}
Two mid-infrared (24\,\micron) observations of 
comet \cg\ and its projected orbit were carried out with the Multiband
Imaging Photometer \citep[MIPS;][]{rieke-young2004} on board the Spitzer Space
Telescope \citep{werner-roellig2004} on 28-29 August 2005 and 8-9 April 2006, respectively.
During the first observation, \cg\ was at a
heliocentric distance of 5.69~AU (out-bound) and at a distance from Spitzer of
5.72~AU. The orbit 
was mapped on a length of 46\arcmin\ projected to the image plane, of which
28\arcmin\ were in the direction trailing the nucleus.
The observation in April 2006 took place at 5.66~AU
from the Sun (in-bound) and 5.26~AU from Spitzer. 
The projected length of the covered orbit section was 42\arcmin,
of which 33\arcmin\ were in the trailing direction.

Each observation was implemented as a cluster in photometry mode:
A set of neighbouring pointing coordinates along the trail was
specified. For each of these, a sequence of exposures ({\it Data
  Collection Events} or DCEs) was taken at 
eight scan mirror positions and repeated with a small offset of the spacecraft
perpendicular to the direction of motion of the scan mirror.
Around each pointing coordinate, the total field of view covered 
was 8.25\arcmin\ $\times$ 8.25\arcmin, while a single DCE covered about
5.25\arcmin\ $\times$ 5.25\arcmin.
The first DCE at each spacecraft position had a
shorter exposure time and was depressed in response. These images were
not used for further analyses, leaving 14 valid exposures at
each specified pointing coordinate.

In August 2005, the set of evaluated data consists of 
252 images of 30\,s exposure time and each covering a field of view of 5.25\arcmin\ $\times$ 5.25\arcmin.
Nine sets of pointing coordinates were
specified along the comet orbit, each of which was covered twice, going from
east to west and back again. Each section of the orbit -- unless close to the edges of the covered area --
was exposed between 28 and 42 times, corresponding to 14 to 21 minutes in total. The images used in the following were
processed with version 16.1.0 of the MIPS pipeline.

In April 2006, eight pointing positions along the orbit were
specified, going three times from east to west and back.
The four easternmost pointing positions were covered only five times.
The resulting data set consists of 672 images of 10 s exposure
time and a field of view of 5.25\arcmin\ $\times$ 5.25\arcmin. The total
exposure 
time for a given area of sky along the orbit varies between 9.8 and 19.8
minutes, and we used the data processed with version 13.2.0 of the MIPS pipeline.

From each exposure, two images were down-linked from the telescope: the 
first difference image, corresponding to an exposure of about 0.5\,s, and
the slope image, representing the flux averaged over
the entire exposure time. 
The images were flatfielded and calibrated in units of MJy/sr by the automated
pipeline at the Spitzer Science Center (SSC). 
A Basic Calibrated Data (BCD) image generated by the SSC pipeline is
normally derived from the slope 
image. If, however, a pixel in the slope image is close to saturation, its
value is replaced by that in the first difference image, in order to determine
the flux from bright sources more reliably.
The slope images from 2005 were close to saturation
because of the long exposure time (30\,s) and the bright background.
A considerable fraction of the pixels in the BCD images 
had therefore been derived from the first difference images, such that
the BCD images largely corresponded to images of 0.5\,s exposure time instead
of 30\,s with 
accordingly lower signal-to-noise ratio (SNR). 
These
images were not suitable for the analysis of the faint trail.
On the other hand, most of the pixels in the slope 
images were not hard saturated, and --
as a part of the calibration pipeline -- the slope images had been
corrected for the nonlinearity of the detector response at high exposure levels. 
Therefore we used the basic calibrated {\em slope} images for the following
evaluation. 

In 2006, the exposure time of an individual image was shorter than in 2005
(10\,s), and the background flux was lower.
Hence, the slope images were not
saturated and served as basis for the BCD images. 
Our analysis of the 2006 observation is based on the BCD images.

To better separate the cometary signal from the background interstellar
medium, and -- to a limited extent -- also from stars and galaxies, shadow
observations were carried out for both Spitzer observations, 
i.\,e.
exact repetitions of the respective primary 
at a later time, after the comet had left
the field of view. 

For technical reasons, the shadow
for 2005 was not observed until April 2007,
and the field of view covered by the shadow was different from that of the
primary.
No shadow is available for the trail at more than
20\arcmin\ distance from the nucleus. 
Since primary and shadow were made from different positions along the
orbit of Spitzer, the zodiacal
light background was different and
had to be removed from the images before the shadow was subtracted
from the primary. 
The shadow for the 2006 observation was taken a week after the primary (14-15
April).

The slope images obtained in 2005 were characterised by (1) a brightness gradient roughly
 from south-west to north-east,
and (2) by a more intricate pattern of
 brightness variation that was common to all images.
The amplitude of the pattern (2) diminished with increasing DCE number. 
\cite{reach-kelley2007} remove both artefacts by 
 subtracting a bias image from each exposure, obtained by
 median-averaging over all exposures of a given DCE number. 
For our data, different background levels in exposures taken
 at the same scan mirror position (same DCE number) precluded the generation
 of a median image in one step. 
For images of given DCE number, the background 
brightness in the primary images increases by about 0.8~MJy/sr
from the easternmost to the westernmost images. 
This correlates well in amplitude and direction with the zodiacal light 
background predicted by the Spitzer Planning Observations Tool (SPOT). The underlying model is based on data obtained with IRAS and with the {\em Cosmic Background
  Explorer} (COBE) and takes into account the dependence of the zodiacal
light background on the position of the telescope.
The predicted contribution of the interstellar
 medium to the background is fainter by an order of magnitude. 

In both primary and shadow of the 2005 observation,
we removed the zodiacal light and the instrumental bias by the following two
steps. We first fitted a plane surface to the zodiacal brightness estimated by
SPOT as a function of the pointing coordinates. We did this independently for the
first and second series of each set of exposures to account for the time
dependence of the zodiacal light due to the changing position of the
spacecraft (about 0.1 MJy/sr over 3 hours).
The obtained plane surface was
subtracted from each frame. The resulting frames had approximately constant
background brightness, allowing us to obtain their median. 

We extracted the
bias image for each DCE number by median averaging frames of the same DCE
number. For the primary, we used only frames located to the east of the comet,
because there is no
detectable trail signal in this part of the FOV (i.e. preceding the nucleus in
its direction of motion). For the shadow we used all frames. The bias images
were subtracted from the respective frames. 

The shadow images still showed a trend of decreasing mean brightness with
time (about 0.3 MJy/sr over 3 hours) which would have led to
noticeable artefacts after averaging. Hence we fitted a linear function
to the image midpoint as a function of the exposure sequence index and
subtracted the respective value from each image. Thus the average background in the shadow was zero and its
subtraction from the primary did not change the background level of the latter,
resulting in a smooth transition between the shadowed and the unshadowed
part (Fig.~\ref{fig:s06_images}).

Solar system objects are not affected by the 
subtraction of shadow from primary, because they have no
counterparts at the same position in the respective other image. 
They appear either  white or black in the
difference image, depending on whether
they were present in the primary or in the shadow.
Solar system objects were excluded from the final mosaic making use of 
their apparent motion between consecutive exposure series:
For each observation, we co-added frames from the different
 exposure series separately (i.e. two for 2005, and six for 2006). 
By visual inspection we identified objects
the position of which changed relative to the stars, and created mask
files to flag these moving objects during the co-addition of all exposures. 
All solar system objects we investigated had a non-zero
apparent speed relative to the comet, thus
none can be considered a candidate for a fragment of \cg. We derive a lower size limit for detectable fragments by comparison with the SNR of the nucleus, which has a radius of 2\,km (\cite{tubiana-barrera2008, kelley-wooden2009}) and was detected in our images at roughly 25$\sigma$. If fragment and nucleus produce dust at the same level per surface area (if any) and have similar thermal properties, a 5$\sigma$ detection would require an object of 0.9\,km radius, corresponding to an absolute R-magnitude of 17.1 (scaled from the absolute magnitude of the nucleus of 15.35 given by \cite{tubiana-barrera2008}).

\begin{figure}[h]
%
\includegraphics[clip,width=\textwidth]{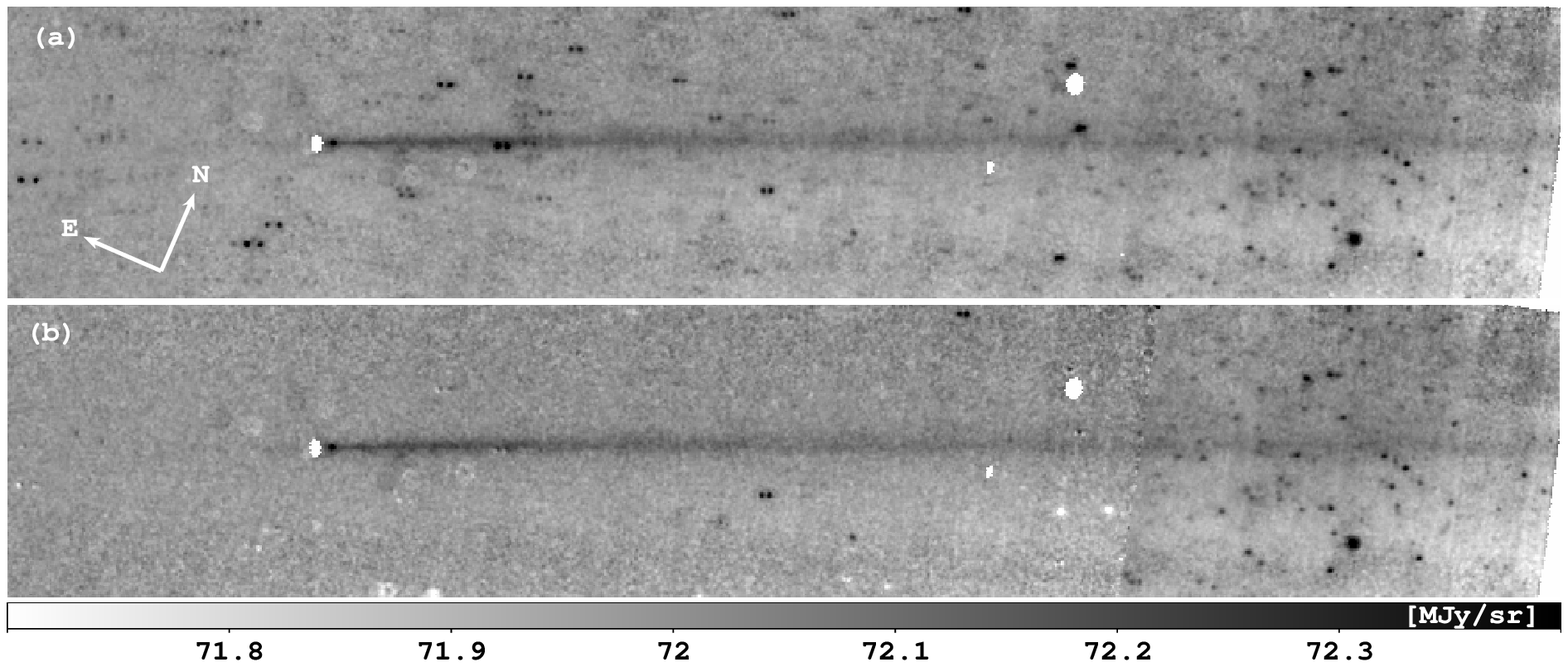}
\includegraphics[clip,width=\textwidth]{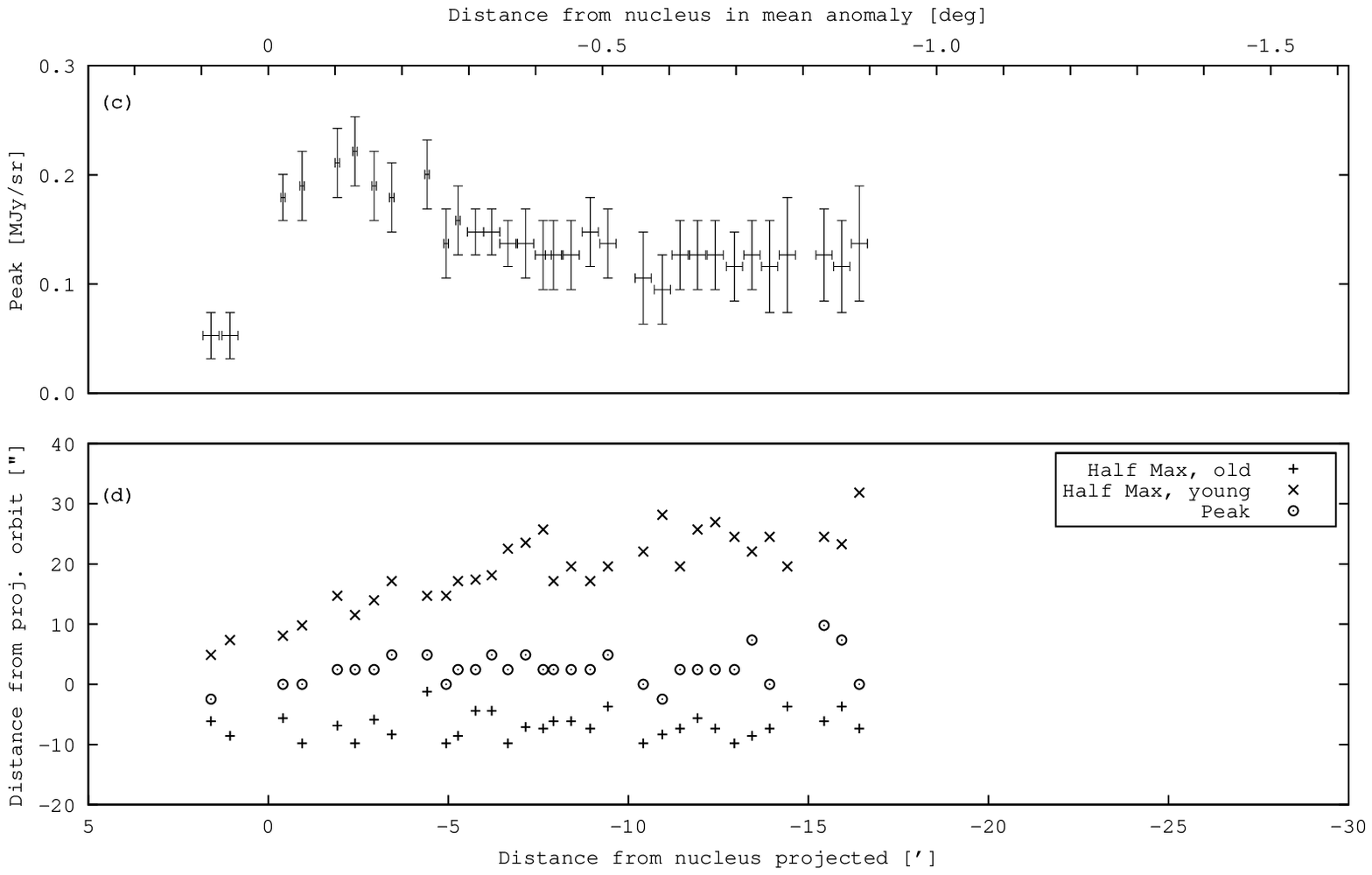}
\caption{Spitzer/MIPS24 observation of the \cg\ trail on 28-29 August 2005.
All panels are aligned in x, the scale in projected distance from the nucleus is given at the bottom of panel d, and the corresponding offset in mean anomaly of the comet is shown at the top of panel c. One arcminute corresponds to about 2.5$\times$10$^8$\,m at the comet. One pixel corresponds to 2.45\arcsec.
{\bf a)} Mosaic of the primary images. The surface brightness is in MJy/sr, and moving objects close
to the comet orbit have been removed.
{\bf b)} The same image after subtraction of the shadow taken in April
2007. 
To derive the data points in panels (c) and (d), we boxcar-filtered the image in panel (b) over 3x3 pixels close to the nucleus and 11x3 further away, and evaluated profiles perpendicular to the projected orbit at discrete distances $x$ from the nucleus. Examples of such profiles are displayed in panels (c) to (f) in Figs.~\ref{fig:best_fit_05}.
{\bf c)} Height above background of the brightest point in the trail cross-section. The error bars in x are the size
of the boxcar averaging window, those in y represent 1$\sigma$ of the 
background variation.
{\bf d)} FWHM of the trail (see caption of Fig.~\ref{fig:wfi_images}).
}
\label{fig:s05_images}
\vspace{10cm}
\end{figure}

\begin{figure}[h]
\includegraphics[clip,width=\textwidth]{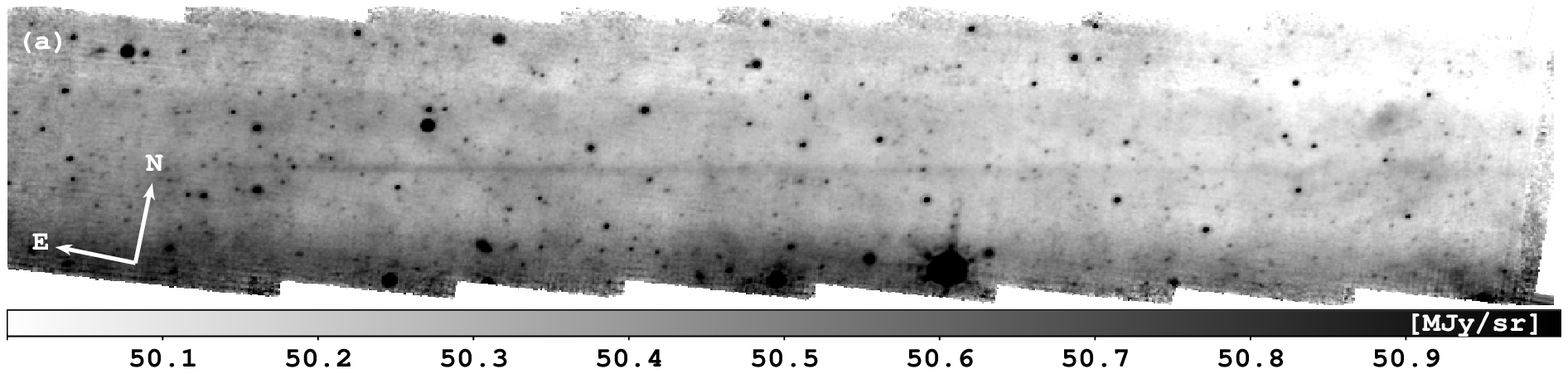}
\vspace{-4mm}

\includegraphics[clip,width=\textwidth]{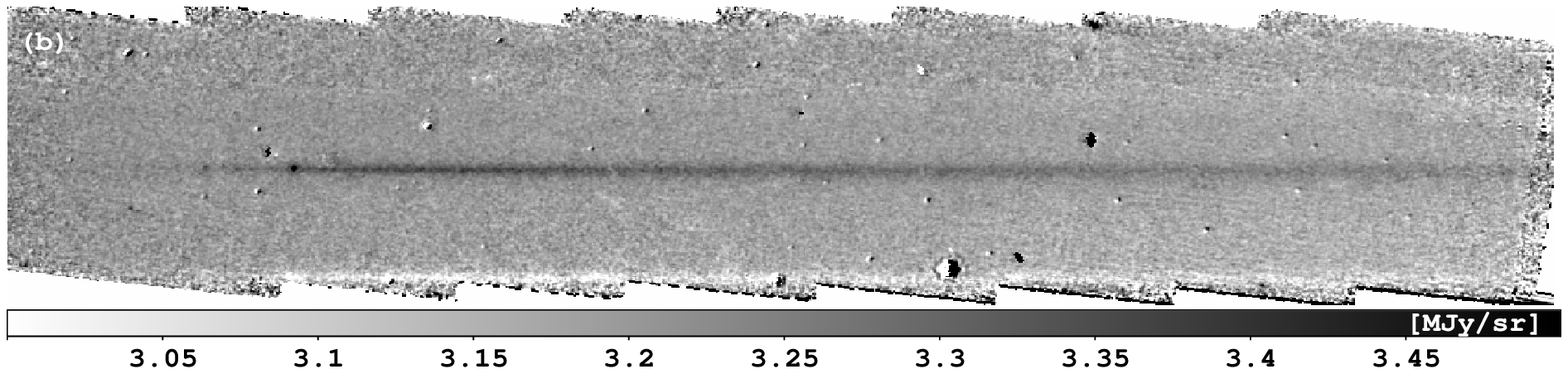}
\includegraphics[clip,width=\textwidth]{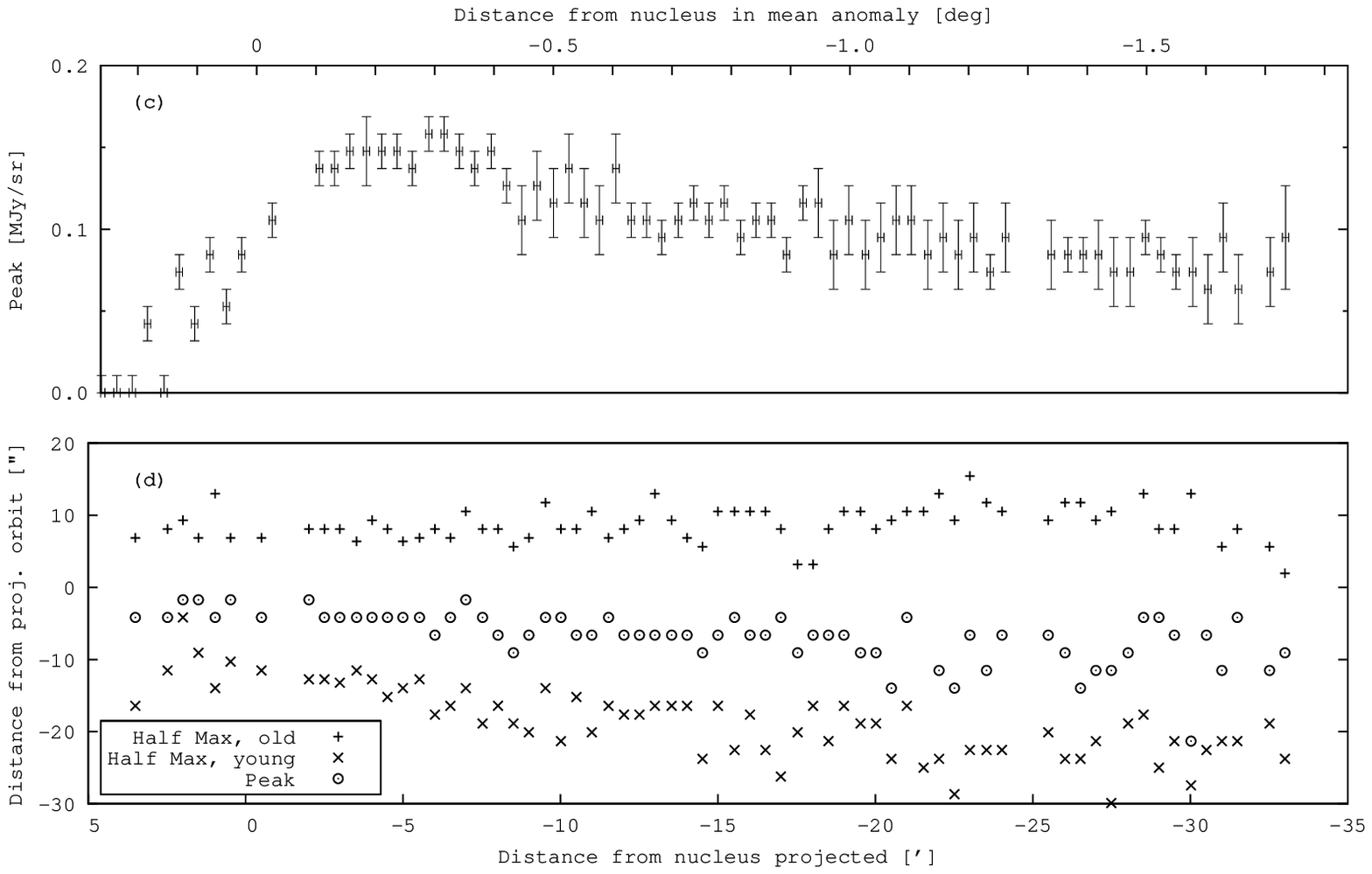}
\caption{Spitzer/MIPS24 observation of the \cg\ trail on 8-9 April 2006.
All panels are aligned in x, the scale in projected distance from the nucleus is given at the bottom of panel d, and the corresponding offset in mean anomaly of the comet is shown at the top of panel c. One arcminute corresponds to about 2.3$\times$10$^8$\,m at the comet. One pixel corresponds to 2.45\arcsec.
{\bf a)} Mosaic of the primary images. The surface brightness is in MJy/sr, and moving objects close
to the comet orbit have been removed.
{\bf b)} The same image after subtraction of the shadow taken six days after the primary. The trail was still in the FOV when the shadow was made and
is visible in white close to the lower margin of the mosaic. 
For {\bf c)} and {\bf d)} see caption of Fig.~\ref{fig:s05_images}. For this image we used a boxcar-filter of size 5x5 pixels.}
\label{fig:s06_images}
\vspace{10cm}
\end{figure}

For the final mosaics we have used the zodiacal-light- and bias-subtracted frames from 2005, and the BCD images from the Spitzer pipeline from 2006, respectively. 
For each mosaic we average-combined all frames for the
primaries and shadows separately in the co-moving frame of the comet in the
primary, masking the moving objects.
Each shadow was subsequently subtracted from the respective primary, and the
results are shown in Figs.~\ref{fig:s05_images} and \ref{fig:s06_images}. 
The trail brightness and width as a function of distance from the
nucleus was measured in perpendicular profiles.
%

The surface brightness of the trail was colour-corrected
to account for the slope of the spectrum of the dust within the
spectral band of MIPS at 24\,\micron:
To derive the true monochromatic flux
density at the weighted average wavelength, $\lambda$ = 23.675\,\micron,
the measured values were 
divided by a colour correction factor of 0.948 (see MIPS Data Handbook) corresponding to colour temperatures between 124\,K and 147\,K (Section~\ref{sec:results}).
The uncertainty of the colour correction factor due to the temperature 
is negligible compared with the variability of the background.

Like the optical image, both Spitzer images show a brightness bulge immediately behind the
nucleus in the sense of its orbital motion, and a comparatively flat profile
further away. Between the three observations, the bulge moved
away from the nucleus and became wider and shallower, both in projected
distance and in terms of mean anomaly, indicating that the dominating
particles have orbital periods longer than that of the nucleus. 
The trail width increases with distance from the nucleus on one flank, and is constant on the other. An interpretation in terms of different
particle sizes and emission times is discussed at the beginning of Sec.~\ref{sec:results}.

The images of the nucleus from both Spitzer observations have central peaks with a full-width at half-maximum (FWHM) of about 3 pixels (7.4\arcsec\ or about 29000\,km) and a surrounding diffraction ring. They are similar to other point sources in the same field of view, thus we do not detect any cometary activity at the times of our observations. However, as pointed out by \cite{kelley-wooden2009}, we cannot exclude the presence of an unresolved coma within the central 29000\,km.

\section{Simulation of Trail Images}
\label{sec:model}
We outline in Section~\ref{subsec:model_outline} the method used to
simulate images of the comet trail; the detailed
mathematics are given in Appendix~\ref{app:math}.
Our physical model of the comet and
the emitted dust is described in Section~\ref{subsec:model_physical}.

\subsection{Simulation Method}
\label{subsec:model_outline}
The position of a dust particle in a cometary trail is a function of 
the emission time and velocity, and of the
radiation pressure coefficient $\beta$, which is the ratio of the solar gravity and radiation pressure on the particle (\cite{burns-lamy1979}). In the
following we assume that the dust emission from
the nucleus is isotropic, that dust trajectories are determined by solar
gravity and radiation pressure only, and that the dust properties do not change
after emission from the nucleus. 
%
A ``dust shell'' describes 
the ensemble of locations occupied by dust particles characterised by a certain
radiation pressure coefficient $\beta$, 
and released at the time $\te$ from the comet in all directions with the
emission speed $\ve$. The time evolution of the shape and size of a dust
shell is described as follows (e.g. \cite{mueller-green2001}):

During a short interval after emission, the shell can be
approximated by a sphere, the centre of which
is off-set from the comet due to radiation pressure
\citep{finson-probstein1968a}. 
The position $\vr$ of a particle in the shell is given by 
\begin{equation}
\vr(t,\beta,\vve) = \vr(t,\beta,0) + (t-t_{\rm e}) * \vve,
\label{eq:spher_shell}
\end{equation}
where $\vve$ is the emission velocity of the
particle, and $\vr(t,\beta,0)$ is the position at the time $t$ of an auxiliary
particle having the same $\beta$ and $t_{\rm e}$ but zero ejection speed. 
The auxiliary
particle does therefore not belong to the ensemble of shell particles.

On longer timescales, the shell is
distorted into an ellipsoid by tidal forces \citep{kimura-liu1977}.
Equation~\ref{eq:spher_shell} was therefore generalised by 
\citet{fertig-schwehm1984} to 
\begin{equation}
\vr(t,\beta,\vve) = \vr(t,\beta,0) + {\mathbf \Phi}(t,t_{\rm e})*\vve.
\label{eq:mass_shell}
\end{equation}
Analytical expressions for the elements of the matrix
${\mathbf \Phi}$ were derived by \cite{massonne1985a, massonnePhD} from Keplerian dynamics.

When the shell becomes more elongated, also the description by an
ellipsoid ceases to be valid. Large particles are then found on a tube bending
around the comet orbit, while small particles have been dispersed by
radiation pressure.
To approximate particle positions on the elongated shell, a local
linearisation was introduced by \cite{mueller-green2001}.
For this purpose, the tube is cut into slices roughly perpendicular 
to its central axis.
The auxiliary particle used in Eqs.~\ref{eq:spher_shell} and
\ref{eq:mass_shell} is replaced by a reference particle
in the centre of the slice \citep{kondrateva-reznikov1985, mueller-green2001},
characterised by the same $\beta$ and $\te$ as the shell, but different $\ve$. 
The position of a particle on the slice $i$ is expressed as 
\begin{equation}
\label{eq:mueller_shell}
\vr(t,\beta,\vve) = \vrref(t) + \Phii(t,t_{\rm e},\beta,\vvref)*(\vve-\vvref), 
\end{equation}
where $\vrref$ is the position
of the reference particle at the observing time and $\vvref$ is its ejection
velocity. 
For Keplerian motion, the
elements of the matrix $\Phii$ can be calculated analytically in analogy to those given by
\citet{massonne1985a, massonnePhD} for the matrix in Eq.~\ref{eq:mass_shell}. 

We use reference particles that were emitted
parallel or anti-parallel to the direction of
the heliocentric velocity of the comet, because
the position of a particle along the
shell is controlled by its orbital period (\cite{mueller-green2001}).
To first order the orbital period of the particle 
depends only on the component of the ejection velocity that was
parallel to the orbital velocity of the comet at the ejection time. 
Particles emitted parallel to the motion of the nucleus will eventually fall back behind it, while particles emitted backwards can overtake it if their emission speed was sufficiently large to compensate for radiation pressure.

We choose the slices through the model tube such that they correspond to pixel
columns in the
simulated image. 
For every $i$, a reference particle
(characterised by $\beta$ and emitted at $\te$)
is determined that is located in the centre of the column 
at the observing time $t$.
For this column, the elements of the matrix $\Phii$ are 
calculated. By inverting Eq.~\ref{eq:mueller_shell}, the emission
direction of a particle 
observed at a given point in column $i$ is obtained. Repeating this
operation for the four corners
of a given pixel, the corresponding fraction of solid angle in the space of
emission directions is calculated. 
The amount of dust located in this pixel follows 
from the overall dust production rate.
By applying the described procedure to all pixels in all columns, the image of one particle shell is
obtained. The final image is constructed as the sum of images of shells 
emitted at various times $\te$ and characterised by various radiation pressure
parameters $\beta$ and emission speeds $\ve$.

The trail observed along the projected orbit of
67P/C--G consists of particles emitted after the close encounter of 67P/C--G
with Jupiter in 1959, which form a narrowly confined stream to date, while
older particles 
were scattered into orbits very different from that of the comet
(J.~Vaubaillon, private communication). 
The dynamics of
these particles {\em relative to the comet} can be approximated by Keplerian
dynamics, while the calculation of an ephemeris of either comet or
dust would require to take into account planetary perturbations. Since we are
only interested in the relative positions, we use the analytical 
expressions for the matrix elements based on the orbital elements of the comet.

The relative position of the nucleus and the dust emitted during previous apparitions is affected by a non-gravitational rocket force due to outgassing (\cite{marsden1969, marsden-sekanina1973}). This force affects the nucleus but not the dust, unless the dust was initially also outgassing. To assess the effect of the rocket force on the nucleus we compared ephemerides of the comet for the dates of our observations with and without the rocket force starting from osculating elements of the comet at point in time between 1959 and 2002. The difference between the predicted positions with and without rocket force gives us a measure of the error in the relative position the nucleus and dust emitted at the epoch of the employed osculating elements. We found that the trail width is affected by less than 2\arcsec, which is negligible for the purpose of our modelling.
The error of the nucleus position along the orbit is $<$5\arcsec\ compared to dust from 2002, $<$14\arcsec\ compared to dust from 1996, increasing up to 130\arcsec\ for dust from 1962. We expect that the distribution of dust from past apparitions is fairly uniform on such spatial scales (cf. Figs.~\ref{fig:best_fit_04} to \ref{fig:best_fit_06}, long-dashed lines), such that the shift due to the rocket force would not noticeably affect the model images.
We therefore have not included the non-gravitational force on the nucleus in our model. 
We have here assumed that the trail particles have not been subject to the rocket force themselves, i.e. have had an extremely low ice content at their emission from the coma. \cite{reach-vaubaillion2009} have inferred this to be the case for the debris trail particles of comet 73P/Schwassmann-Wachmann~3 from their proximity to the nucleus' orbit.
If, however, the trail particles had any appreciable covering of ice, it would indeed have affected their trajectories significantly.

\subsection{Physical Model of the Cometary Emission}
\label{subsec:model_physical}

%
In this section, we describe the physical model of the cometary emission on which our results are based. We declare, which quantities are input to our model, which are output, and which cannot be constrained with our method. 

In our model, dust is emitted isotropically from the nucleus. Images of the coma have shown azimuthal brightness variations during the 1996/97 and 2002/03 apparitions (\cite{schulz-stuewe2004a,
  schulz-stuewe2004b, weiler-rauer2004a, schleicher2006}). But no quantitative model of the distribution of activity across the surface of comet \cg\ has been published to date, such that isotropic emission remains the simplest possible assumption. 
The dust emission in our model does not change between different apparitions
of the comet after the last close encounter with Jupiter in 1959, because we
consider the available data on gas production and \afr\ too scarce to
support a more elaborate model.
The comet in our model is active inside 3.1 AU, corresponding to a time interval of 600 days around perihelion. 
We consider discrete particle classes,
characterised by a specific value of the radiation pressure coefficient $\beta_i$, and offset from each other by a constant factor in $\beta$.

Our observations were made at 4.7 AU from the Sun and further, i.e. the first observation was made 10 months after the assumed cessation of activity. Hence, small particles with high sensitivity to radiation pressure ($\beta>0.1$ in 2004 and $\beta>0.01$ in 2005/06) will have left the FOVs of our observations. For bulk densities of 2000\,kg/m$^3$ and lower, all particles in the FOVs are larger than the wavelengths of the observed light.

To simulate the optical observation, we assume that the light back-scattered
by the dust has the same spectral properties as sunlight, i.e. that the dust
albedo is independent of (visible light) wavelength and particle size. In a recent paper \cite{tubiana-barrera2008} have found that the trail dust has a red spectral colour, but since our observation was made without filter, we can only derive the average albedo in the visible wavelength range. 
The intensity incident on the particle is 
given by $J_{\rm inc}$ = $\Isun / r_{\rm h}^2$, with the solar flux at 1\,AU
$\Isun$ = 1367 W/m$^2$, and the heliocentric distance of the particle
$r_{\rm h}$ in AU. We apply the geometric optics approximation, 
because the particles in the FOV are expected to be larger than
visible wavelengths.
The distance between particle and observer is $\Delta$ (in m). We denote the total cross-section of particles characterised by $\beta_i$ in a given pixel by $S_i$, i.e. $S_i = N_i \pi s_i^2$, where $N_i$ is the number of particles and $s_i$ the particle size. Our method helps to constrain $S_i$, but neither $N_i$ nor $s_i$ (see below).
The intensity received by the detector from all particles in the pixel is 
\begin{equation}
J_{\rm rec}^{\rm vis} = \frac{p \, j(\alpha)}{\pi} \frac{\Isun}{r_{\rm h}^2} 
\frac{1}{\Delta^2} \sum_i S_i, 
\end{equation}
where $p$ is the geometric albedo and $j(\alpha)$ the phase function.

The Spitzer images are simulated using the monochromatic flux of a blackbody 
at the weighted average wavelength of MIPS24
(23.675\,\micron). 
The intensity per unit frequency interval $J_{\rm rec}^{\rm IR}$  received by
a detector pixel at a distance $\Delta$ from the particles having $\beta_i$ is described by
\begin{equation}
J_{\rm rec}^{\rm IR} = \frac{1}{\Delta^2} \epsilon B_{\nu} (\nu, T) \sum_i S_i,
\label{eq:fluxpp}
\end{equation}
where $B_{\nu} (\nu, T)$ is Planck's function, $\nu$ is the frequency, $\epsilon$ the emissivity at 24\,\micron, and $T$ the temperature.
The flux $J_{\rm rec}^{\rm IR}$ in Eq.~\ref{eq:fluxpp} is converted to units of MJy/sr by
the factor $10^{20}/\df \omega$\,(pixel), where
$\df \omega$\,(pixel) is the solid angle covered by the pixel.

Since our observations are at a single infrared wavelength, we cannot infer the
dust temperature from them. The temperature 
determines the intensity of the blackbody radiation for a given cross-section via Planck's function. With higher temperature, less cross-section is required in the FOV to achieve the same infrared flux.
Assuming that the particles are characterised by the Bond albedo $A_{\rm B}$ at
visible wavelengths and the emissivity $\epsilon$ at 24\,\micron, their temperature $T$ at the heliocentric distance $r_{\rm h}$ 
results from the equilibrium between absorbed solar and emitted thermal
radiation:
\begin{equation}
T (r_{\rm h}, A_{\rm B},\epsilon)
= 278.8 \,{\rm K} \, \left(\frac{1-A_{\rm B}}{\epsilon}\right)^{\!\frac{1}{4}} \, \frac{1}{\sqrt{r_{\rm h}/{\rm AU}}}.
\label{eq:T_eq}
\end{equation}
\cite{sykes-walker1992a} derived $\epsilon/(1-A_{\rm B}) = 0.6 \pm 0.2$ from IRAS measurements of the trail brightness at three wavelengths (12, 25, and 60\,\micron), which at the heliocentric distance $r_{\rm h}$=5.7 AU during our observations corresponds to temperatures between 124K and 147K. The IRAS-derived colour temperatures are higher than the temperature of a blackbody ($\epsilon$=1 and $A_{\rm B}$=0) at the same heliocentric distance. The observed excess colour temperatures are consistent with randomly oriented, rapidly rotating, large, and dark particles capable of sustaining a latitudinal temperature gradient (\cite{sykes-gruen-cometsII-2004}).

The emission speed of a particle from the coma depends to first order on its cross-section-to-mass ratio. We denote the inverse cross-section-to-mass ratio by $\chi$, i.e. for spherical grains of the radius $s$ and the bulk density $\rho$, $\chi = \rho s$.
Hydrodynamic models suggest that $v \propto (\chi)^{-0.5}$ (e.g. \cite{crifo-rodionov1999a, agarwal-mueller2007}). We use the expression
\begin{equation}
v\,(\chi,r_{\rm h}) = f_{\rm v} v_0 \left(\!\frac{\chi}{\chi_0}\!\right)^{\!\!-0.5} 
\!\left(\frac{r_{\rm h}}{r_{\rm p}}\!\right)^{\!\!-3},
\label{eq:em_speeds_ana}
\end{equation}
with $v_0$ = 3.9\,m/s, $\chi_0$ = 1 kg/m$^2$, and the
perihelion distance $r_{\rm p}$ = 1.29 AU. For the dependence on the heliocentric distance, a power-law with the exponent of -3 is assumed, because the emission speeds are to first order proportional to the square root of the gas production rate, and the observed water production rate can be approximated by a power-law with an exponent of about -6 (cf. \cite{agarwal-mueller2007} and references therein).
The factor $f_{\rm v}$ is a variable parameter. For $f_{\rm v} = 1$, the
emission speeds at perihelion correspond to those predicted by a hydrodynamic
coma model (\cite{landgraf-mueller1999}) for a perihelion gas
production rate  of $Q_{\rm gas}$ = 10$^{28}$  molecules/s (\cite{agarwal-mueller2007}). 
If Eq.~\ref{eq:em_speeds_ana} yields a value exceeding the speed of
the gas in the cited isotropic coma model, it is set to $v$ = $v_{\rm
gas}$ = 750 m/s. (The decrease in gas speed with heliocentric distance
  is small 
compared with that of the dust speeds and is not taken into account.)
If the speed according to Eq.~\ref{eq:em_speeds_ana} is below the escape
speed $v_{\rm esc}$, the particle is not further considered. We use $v_{\rm esc}$=0.2 m/s, which corresponds to the escape speed at 20 km distance from the centre of a nucleus having a mass of 8 $\times$ 10$^{12}$ kg (\cite{lamy-toth2006, lamy-toth2007}) and radius of 2\,km (\cite{tubiana-barrera2008, kelley-wooden2009}).

The radiation pressure coefficient $\beta$ is characterised by the radiation pressure efficiency
$Q_{\rm pr}$, and the inverse cross-section-to-mass ratio $\chi$:
\begin{equation}
\beta = 
\frac {3 \,L_\odot}{16 \,\pi c\, G M_\odot} \frac{Q_{\rm pr}}{\chi}
= k_{\beta} \frac{Q_{\rm pr}}{\chi},
\label{eq:beta_model}
\end{equation}
where $L_\odot$ and $M_\odot$ are the luminosity and mass of the
Sun,  and  $k_{\beta}$ = 5.77 $\times$ 10$^{-4}$ kg/m$^2$ (\cite{burns-lamy1979}). 

%
Both Eqs.~\ref{eq:em_speeds_ana} and \ref{eq:beta_model} depend on $\chi$. Since $v$ and $\beta$ are in our model the only quantities determining the dynamics of a particle, each particle can only be characterised by its value of $\chi$, not by its size, because we have no means to infer the bulk density $\rho$ of the dust particles.
Equations~\ref{eq:em_speeds_ana} and \ref{eq:beta_model} can be combined to 
\begin{equation}
v(\beta, r_{\rm h}) = \frac{f_{\rm v}}{\sqrt{Q_{\rm pr}}} v_0 \sqrt{\frac{\chi_0}{k_\beta}} \!\left(\frac{r_{\rm h}}{r_{\rm p}}\!\right)^{\!\!-3} \sqrt{\beta}.
\label{eq:v_of_beta}
\end{equation}
The variable parameter in this equation is $f_{\rm v}/\sqrt{Q_{\rm pr}}$. 
%

Our method allows us to constrain the total cross-section of particles characterised by a given value of $\beta$ emitted by the comet as a function of time. For constant bulk density $\rho$, the relative contribution to the total cross-section of particles in an interval $(\beta, \beta+\df \beta)$ is described by the distribution function $f(\beta)$ defined by \cite{finson-probstein1968a}:
\begin{equation}
\chi^2 g(\chi) \df\chi \propto f(\beta) \df \beta,
\label{eq:sd}
\end{equation}
where $g(\chi) \df\chi$ is the number of particles in the interval $(\chi,\chi
+ \df \chi)$ and $\beta(\chi)$ is given by Eq.~\ref{eq:beta_model}. For constant $\rho$, $g(\chi)$ corresponds to a differential size distribution. We represent $f(\beta)$ by a power-law $f(\beta) \propto \beta^a$. For constant $\rho$, the corresponding differential size distribution function is also a power-law $g(\chi) \propto \chi^\alpha \propto s^\alpha$ with the exponent $\alpha = -a - 4$. The exponent $a$ is a free parameter in our study. 

We simulate images for a range of values of $f_{\rm v}/\sqrt{Q_{\rm pr}}$ and
of $a$. From the similarity between simulation and observation, we determine the most appropriate set of values for $f_{\rm v}/\sqrt{Q_{\rm pr}}$ and $a$.

We describe the cross-section production rate of particles in the interval ($\beta, \beta+\df \beta$) for constant $\rho$ by 
\begin{equation}
Q_{{\rm S}}(\beta,\df \beta, r_{\rm h}) = Q_0 \, f(\beta) \, \df \beta \, r_{\rm h}^{-8}. 
\label{eq:Qs}
\end{equation}
$Q_0$ is a free parameter that we constrain by matching the brightness of the simulated trail to the corresponding observation. 
The dependence on heliocentric distance $r_{\rm h}^{-8}$ corresponds for an isotropic coma in steady state to an \afr\ parameter (\cite{ahearn-schleicher1984}) proportional to $r_{\rm h}^{-5}$, which is in agreement with a power-law fit to \afr\ data measured during several apparitions of the comet (\cite{agarwal-mueller2007} and references therein). 

For discrete particle classes and a given relation $s_i (\beta_i)$, we have $Q_{{\rm S},i}(\beta_i) = Q_i \pi s_i^2(\beta_i)$, where $Q_i$ is the number production rate of particles of size $s_i$. Not being in a position to constrain $s_i (\beta_i)$, we cannot infer $Q_i$. We can, however, calculate the mass production rates $Q_{{\rm m},i}$ of the dust for a given value of $Q_{\rm pr}$: 
\begin{equation}
Q_{{\rm m},i}= \frac{4}{3} \,k_\beta \, \frac{Q_{\rm pr}}{\beta_i} \, Q_{{\rm S},i}.
\label{eq:Qm}
\end{equation}
 
From comparison of the cross section production rates required to reproduce the optical
and infrared images, respectively, we infer the visible light albedo, because the
brightness of dust at optical wavelengths is proportional to the albedo, while
the thermal emission is to first order independent of it. 
Since the optical and the first infrared observation were separated in time by
16 months, we can only derive the albedo on the assumption that our model
correctly reproduces the temporal evolution of the trail.

We have simulated images with the original pixel scale of the corresponding observation (0.71\arcsec/pixel for WFI and 2.45\arcsec/pixel for Spitzer). For WFI, we have added Gaussian noise with the parameters characterising the background of the observation, and simulated the flatfielding process and the subsequent spatial filtering. The Spitzer simulations were convolved with a Gaussian having $\sigma=2.3$\arcsec\ to simulate the point spread function (PSF), and spatially averaged over the same ranges as the corresponding observation: 3x3 (11x3) pixels in the inner (outer) part of the 2005 image, and 5x5  pixels in 2006.

\begin{table}[h]
\caption{Free parameters of the employed model. Columns 1 -- 3 summarise the physical model outlined in Section~\ref{sec:model}, column 4 describes the main effect(s) a parameters has on the simulated images (see Section~\ref{sec:results}).}
%
\centering
\label{tab:variables_summary}
\begin{tabular}{lllll}
\hline\noalign{\smallskip}
Physical quantity & Model & Variable & Related observational quantity\\
\hline\hline\noalign{\smallskip}
Diff. $\beta$ distribution & 
Eq.~\ref{eq:sd} &
$a$ &
Brightness far behind nucleus\\
&&& relative to peak; Trail width.
\\
Emission speed & 
Eq.~\ref{eq:v_of_beta}& 
$f_v/\sqrt{Q_{\rm pr}}$ & 
Profile close to nucleus \\
&&&(peak position and profile \\
&&&in front of peak); Trail width.
\\
Cross-sec. prod. rate&
Eq.~\ref{eq:Qs}&
$Q_0$&
Absolute brightness of trail \\
&&&in IR for given temperature.
\\
Geometric albedo &
$p \neq p\,(\lambda, s)$&
$p$ &
Relative brightness in IR and visible.
\\
\noalign{\smallskip}\hline
\end{tabular}
\end{table}

\section{Results}
\label{sec:results}
\begin{figure}[t]
\includegraphics[clip,width=\textwidth]{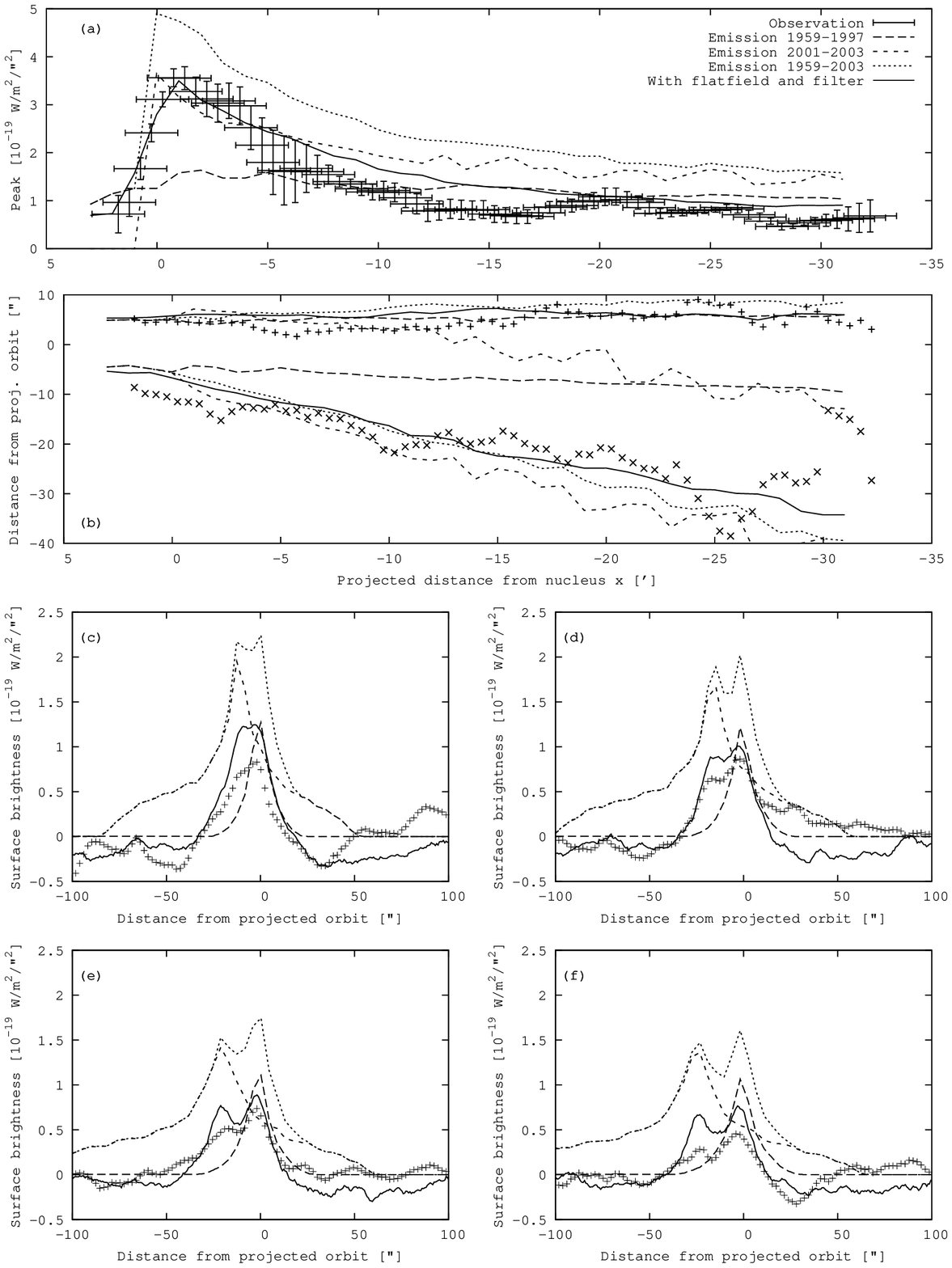}
\caption{
Simulations of the 2004 WFI image with the best-fit parameters $a=0.1$ and $f_{\rm v}/\sqrt{Q_{\rm pr}} = 1.6$. Panels (a) and (b) show the brightness profile and the FWHM as functions of the distance from the nucleus, panels (c) to (f) show brightness profiles perpendicular to the projected orbit at different distances from the nucleus as indicated in the upper right corner of the plots. Separate simulations were run with only particles emitted at least one orbital period before the observation (long-dashed) and emitted during the most recent apparition of the comet (short-dashed). The dotted lines correspond to the sum of both simulations, and the solid lines show the result of including the flatfield and spatial filtering in the simulation. Crosses mark the observation.
}
\label{fig:best_fit_04}
\end{figure}

\begin{figure}[t]
\includegraphics[clip,width=\textwidth]{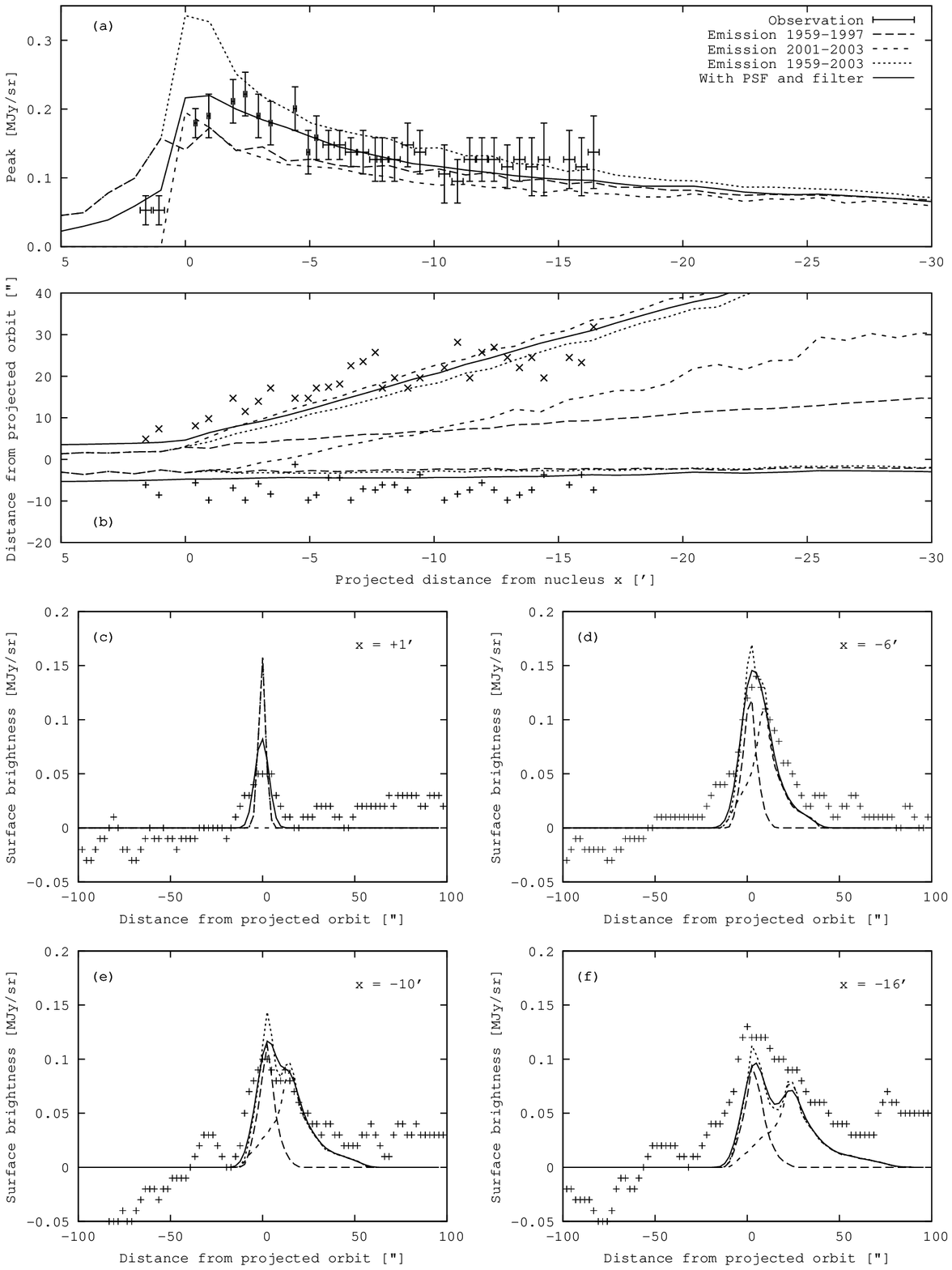}
\caption{Simulations of the 2005 Spitzer image with the best-fit parameters $a=0.1$ and $f_{\rm v}/\sqrt{Q_{\rm pr}} = 1.6$. For details see Fig.~\ref{fig:best_fit_04}. The solid lines show the result including the PSF and spatial filtering in the simulation.
}
\label{fig:best_fit_05}
\end{figure}

\begin{figure}[t]
\includegraphics[clip,width=\textwidth]{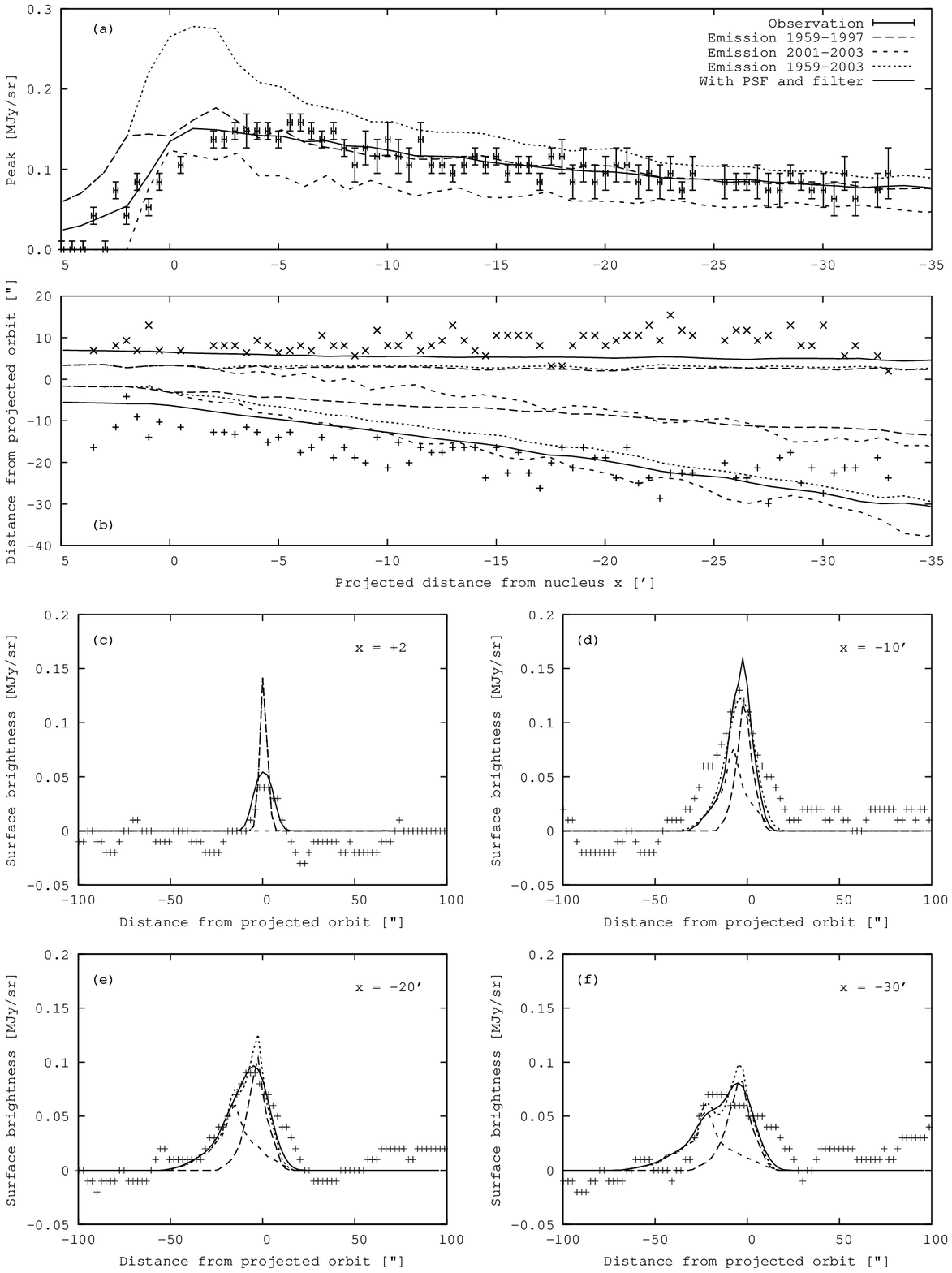}
\caption{Simulations of the 2006 Spitzer image with the best-fit parameters $a=0.1$ and $f_{\rm v}/\sqrt{Q_{\rm pr}} = 1.6$. For details see Fig.~\ref{fig:best_fit_04}. The solid lines show the result including the PSF and spatial filtering in the simulation.
}
\label{fig:best_fit_06}
\end{figure}

We have simulated images for a wide range of parameters. By comparing them 
with the observations we constrain the 
emission speeds of dust particles as a function of the radiation pressure coefficient, the exponent of the dust $\beta$ distribution, the geometric albedo
of the particles, and the dust production rates.
Observations and simulations are compared by evaluating the peak surface
intensity and the FWHM of the trail as functions of distance from the
nucleus in plots like those shown in Figs~\ref{fig:best_fit_04} to \ref{fig:best_fit_06}. We first describe 
the spatial distribution of particles depending on their emission time.
Then we assess the impact of separately varying the speed and $\beta$
distribution parameters, $f_v/\sqrt{Q_{\rm pr}}$ and $a$. We derive the mass production rates and \afr\ parameter for the best-fitting set of parameters. From comparison of the optical and
infrared simulation results, we constrain the particle albedo.  

In all three observations (Figs.~\ref{fig:wfi_images} to \ref{fig:s06_images}), the distance of the half maximum from the projected orbit is approximately constant on one side of the trail and
increases with distance from the nucleus on the other side. The constant side
is marked by ``+'' in the plots, the other by ``x''. 
The ``x''-side appears in projection on the southern side of the trail in the
2004 and 2006 images, and on the northern side in 2005. Defining the ``upper'' side of an orbit as the one out of which
the positive angular momentum vector points, the observer was below the
orbital plane of the comet during the 2004 and 2006 observations (orbit plane angles $-$0.7\degree\ and $-$1.2\degree, see
Table~\ref{tab:obs_geom}) and above the plane in 2005 (orbit plane angle 1.3\degree). Assuming
that all dust remains close to the comet orbital plane, the particles on the
``x''-side are outside the orbit as seen from the inner solar
system. They must therefore be
subject to stronger radiation pressure than those remaining close to the
orbit. 

In Figs.~\ref{fig:best_fit_04} to \ref{fig:best_fit_06} the best-fitting simulations are displayed. Before describing how 
we searched for the best fit, we discuss now the spatial separation of particles emitted during the 2002/03 apparition of the comet on the one hand, and during earlier apparitions on the other. We discuss this on the example of the best-fit simulation, but the geometry is comparable for all other sets of parameters. 

Figures~\ref{fig:best_fit_04} to \ref{fig:best_fit_06} show profiles and FWHM
from four simulations: dust emitted at least one orbital period before the
observation (1959-1997, long-dashed), dust emitted during the apparition
immediately preceding the observation (2002/03, short-dashed), their sum
(dotted), and the result after simulation of the flatfielding for the visible
image, the point spread function for the infrared images, and the
spatial filtering for all (solid). The solid line is the profile that must match the observation.

\begin{figure}[]
\includegraphics[clip,width=\textwidth]{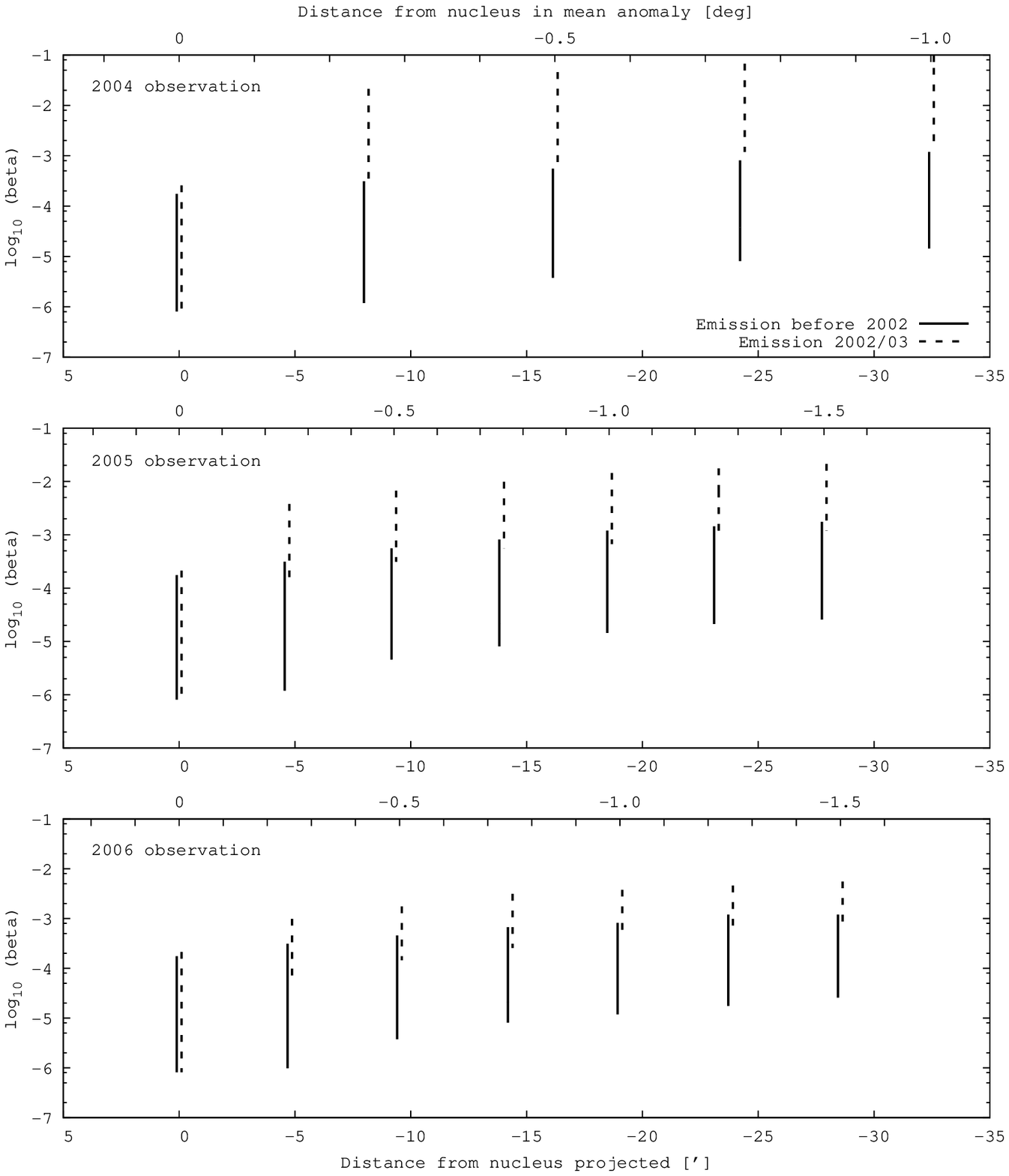}
\caption{Ranges of the radiation pressure coefficient $\beta$ of trail particles as a function of the distance from the nucleus. The $\beta$ ranges were evaluated at multiples of 0.25\degree\ in mean anomaly behind the nucleus in simulations of all three observations with the best-fitting set of parameters ($a=0.1$ and $f_{\rm v}/\sqrt{Q_{\rm pr}}=1.6$). 
The solid lines refer to particles stemming from the apparition immediately preceding the observations (in 2002/03), the dashed lines to particles older than one orbital period (emitted before 2002). Both were evaluated at the same nucleus distance but plotted slightly off-set from each other for better visibility. The lines indicate the presence of particles having a given $\beta$ at the given nucleus distance, not their abundance. The contribution of particles with $\beta>0.01$ to the total cross-section in the 2004 simulation is on the order of 10\%.
}
\label{fig:beta_distance}
\end{figure}

Particles emitted 1959-1997 are concentrated around the projected orbit of the comet. Particles from the 2002/03 apparition form a second profile with a slightly different position angle, which is the cause for the broadening of the trail profile towards the side marked ``x'' in the b-panels of Figs.~\ref{fig:wfi_images} to \ref{fig:s06_images}.
For a given projected distance behind the nucleus (x-coordinate in panels a and b), the 2002/3 particles are more sensitive to radiation pressure than the older particles (Fig.~\ref{fig:beta_distance}).
The difference in position angle between the two profiles is  ($-$0.80 $\pm$ 0.14)\degree\ in 2004,
($+$1.24 $\pm$ 0.15)\degree\ in 2005, and ($-$0.60 $\pm$ 0.15)\degree\ in
2006, where the uncertainty
corresponds to the pixel scale employed for the simulations. 

\input{results_04}
\input{results_05}
\input{results_06}

To investigate how the shape and width of the simulated profiles depend on the emission speed factor  $f_{\rm v}/\sqrt{Q_{\rm pr}}$  and $\beta$ distribution exponent $a$, we have run simulations for all three observations for 105 combinations of $f_{\rm v}/\sqrt{Q_{\rm pr}}$ in the range of 0.2 to 10 and $a$ between -0.4 and 0.4. The absolute brightness in the simulation is controlled by the total cross-section of dust produced, which is a free parameter in our model. The production rates were adjusted in such a way that the peak brightness close to the nucleus matches the observation.
The employed parameter values and the results are described in Tables~\ref{tab:profile_eval_04} to \ref{tab:profile_eval_06}. The tables show a set of five entries for each parameter set. Each entry can be ``+'' (too much), ``O'' (acceptable), or ``--'' (too little). The first three of the five entries refer to parallel profiles like panels (a) in Figs.~\ref{fig:best_fit_04} to \ref{fig:best_fit_06}. (P) is the position of the peak relative to the nucleus, (F) is the brightness to the left of the nucleus, i.e. of dust preceding the nucleus in its orbital motion, (D) is the brightness in the distant trail to the right of the nucleus, trailing it. The two remaining entries in each field refer to the trail width as shown in panels (b) of Figs.~\ref{fig:best_fit_04} to \ref{fig:best_fit_06}. (T) is the width on the flank with the old particles, (N) is the width on the flank with particles from 2002/03. 
 
In the following, we summarise the results shown in
Tables~\ref{tab:profile_eval_04} to \ref{tab:profile_eval_06}.
We first describe how the choice of $f_{\rm v}/\sqrt{Q_{\rm pr}}$ and $a$
affects the parallel profiles of the kind shown in
Figs.~\ref{fig:best_fit_04} to \ref{fig:best_fit_06} (a): 
The factor $f_{\rm v}/\sqrt{Q_{\rm pr}}$ mainly influences the profile of the trail close to the nucleus, i.e. the position of the peak and the shape of the profile in front of it. These characteristics are only little influenced by the choice of the $\beta$ distribution exponent.
When $f_{\rm v}/\sqrt{Q_{\rm pr}}$ is increased, the bright peak close to the nucleus moves further in the direction of the orbital motion of the nucleus, and the brightness in front of the nucleus increases. Particles in front of the nucleus were emitted backwards and then overtook the comet, because due to their smaller total energy they are on orbits with smaller semimajor axis and shorter orbital period. A higher emission speed component anti-parallel to the direction of motion of the nucleus implies a larger difference in orbital period, thus the particles are found further to the left at a given time. Comparison of the long- and short-dashed curves in Figs.~\ref{fig:best_fit_04} to \ref{fig:best_fit_06} (a) reveals that most of the dust in front of the nucleus was emitted more than one orbital period before the observation. The sharp flank left of the nucleus is due to particles from the most recent perihelion passage. 
The slope of this flank is independent of emission speed and $\beta$ distribution, but the position of the peak moves to the left for increasing speed. 

Separate simulations for intervals of the considered ranges of $\beta$ and the emission time have shown that the point of the peak consists of particles emitted backwards after perihelion. All these particles have approximately the same distance from the nucleus. When the model speeds of all particles are increased, at the time of observation all these particles are located more to the direction in front of the nucleus, thus the peak shifts to the left. The particles in front of the nucleus were emitted around perihelion.

The trail profile far behind the nucleus depends mainly on the $\beta$ distribution. The particles found at larger distance from the nucleus are generally smaller than those closer to it, because they fall back more quickly due to radiation pressure. If we increase the relative amount of small particles by choosing a smaller size distribution exponent $\alpha$ (larger $a$), the relative brightness at large distance behind the nucleus increases.

The width of the trail (see Figs.~\ref{fig:best_fit_04} to \ref{fig:best_fit_06} (b)) on both sides of the orbit increases with the emission speed and for $\beta$ distributions favouring smaller particles, the latter because smaller particles have higher speeds perpendicular to the orbit.
 
A well-fitting set of parameters is characterised by a set of five ``O'' symbols in Tables~\ref{tab:profile_eval_04} to \ref{tab:profile_eval_06}  (shaded in grey). For the 2004 WFI image, good fits were obtained with $0.6 \leq f_{\rm v}/\sqrt{Q_{\rm pr}} \leq 2.2$ at $a = -0.2$ and with $1.4 \leq f_{\rm v}/\sqrt{Q_{\rm pr}} \leq 2.4$ at $a = 0$. The Spitzer images can both be fitted with $f_{\rm v}/\sqrt{Q_{\rm pr}} = 1.6$ and $a = 0.2$. There is no set of parameters that is ideal for both the WFI and the Spitzer data. As the best compromise we adopt the values  $f_{\rm v}/\sqrt{Q_{\rm pr}} = 1.6$ and $a = 0.1$. The corresponding simulations are shown in Figs.~\ref{fig:best_fit_04} to \ref{fig:best_fit_06}. 

\begin{figure}[]
\center
\includegraphics[clip,width=.5\textwidth]{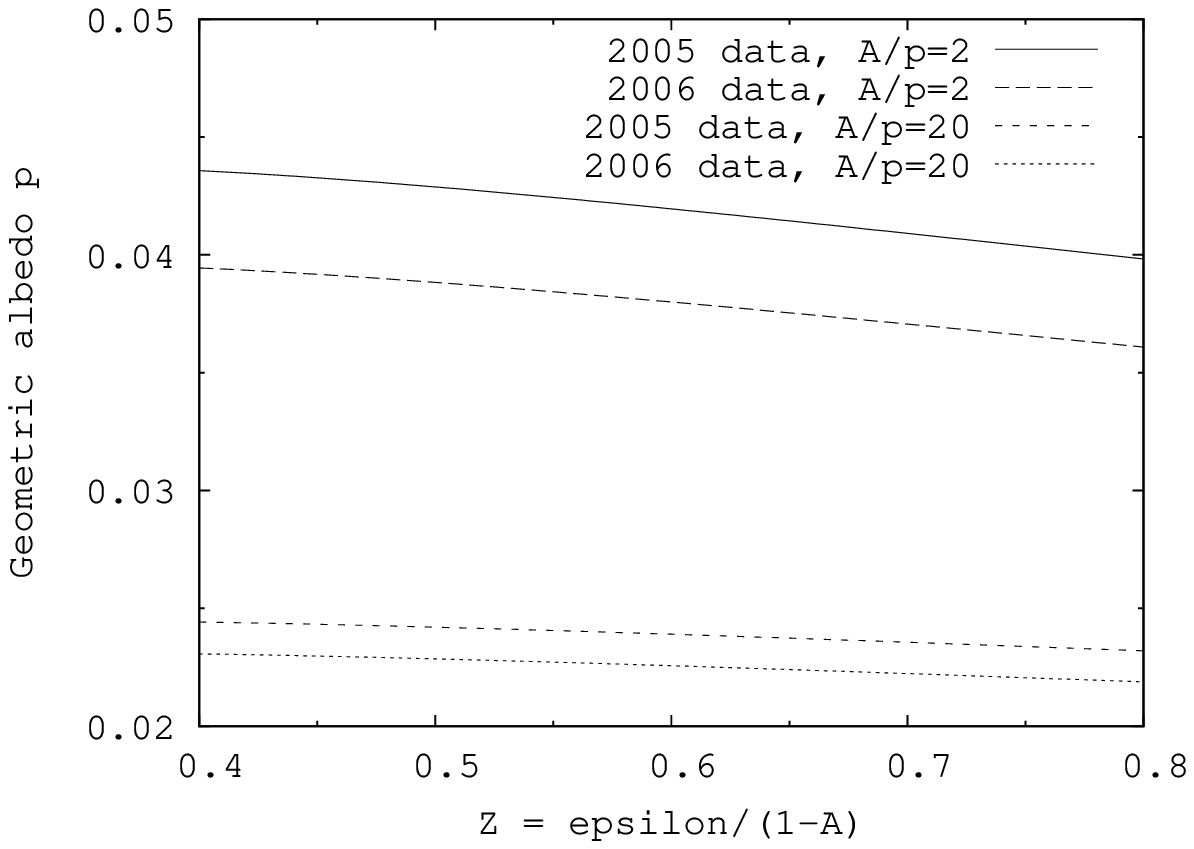}
\caption{Geometric albedo of the dust at visible wavelengths as a function of the parameter $Z=(\epsilon/(1-A))$ in the range of $0.4<Z<0.8$ (\cite{sykes-walker1992a}). The function $p(Z)$ is derived from the condition that the visible and infrared observations must be reproduced with the same dust production rate. We plot $p(Z)$ separately for the 2005 and 2006 observation, and for two extreme values of the phase integral $q=A/p$.
}
\label{fig:albedo}
\end{figure}

By fitting the brightness of the simulated profiles to the observations we
have constrained the product of geometric albedo, phase function, and
cross-section production rate $\tilde{Q_i}^{04} = p j(\alpha) Q_{{\rm S},i}$
from the optical observation. The 2004 observation was made at a
phase angle of 1\degree, and we assume henceforth that $j=1$. From the infrared data, we have constrained the product of emissivity, Planck function, and cross-section production rate $\tilde{Q_i}^{05/06} = \epsilon B_\nu(\nu,T) Q_{{\rm S},i}$. 
We express the visible geometric albedo as a function of the dust temperature and emissivity: 
\begin{equation}
p = \tilde{Q_i}^{04} / \tilde{Q_i}^{05/06} \epsilon  B_\nu(\nu,T),
\end{equation}
which is independent of $i$ because we assume that the albedo and emissivity do not depend on the particle size.
Introducing $Z=\epsilon/(1-A_{\rm B})$ and the phase integral $q=A_{\rm B}/p$
gives $p = \tilde{Q_i}^{04} / \tilde{Q_i}^{05/06} (1-pq) Z B_\nu(\nu,Z)$,
where the relation between $T$ and $Z$ is given by Eq.~\ref{eq:T_eq}. Solving
for $p$ gives the geometric albedo as a function of $Z$. In
Fig.~\ref{fig:albedo} we plot $p(Z)$ derived separately from the 2005 and 2006
observation and for two extreme values of $q$. We have used these extreme
values to get a feeling for the importance of $q$. A common value would be
$q=4.55$, as derived from the phase function given in \cite{divine1981}.
We require a geometric albedo between 0.022 and 0.044 to reproduce both the visible and the infrared data with the same model parameters and in a manner consistent with \cite{sykes-walker1992a}. Similarly low albedo values were observed in dust trails by \cite{ishiguro-watanabe2002, ishiguro-sarugaku2007} and in the coma of comet 1P/Halley by \cite{tokunaga-golisch1986}.

\begin{figure}[]
\center
\includegraphics[clip,width=.5\textwidth]{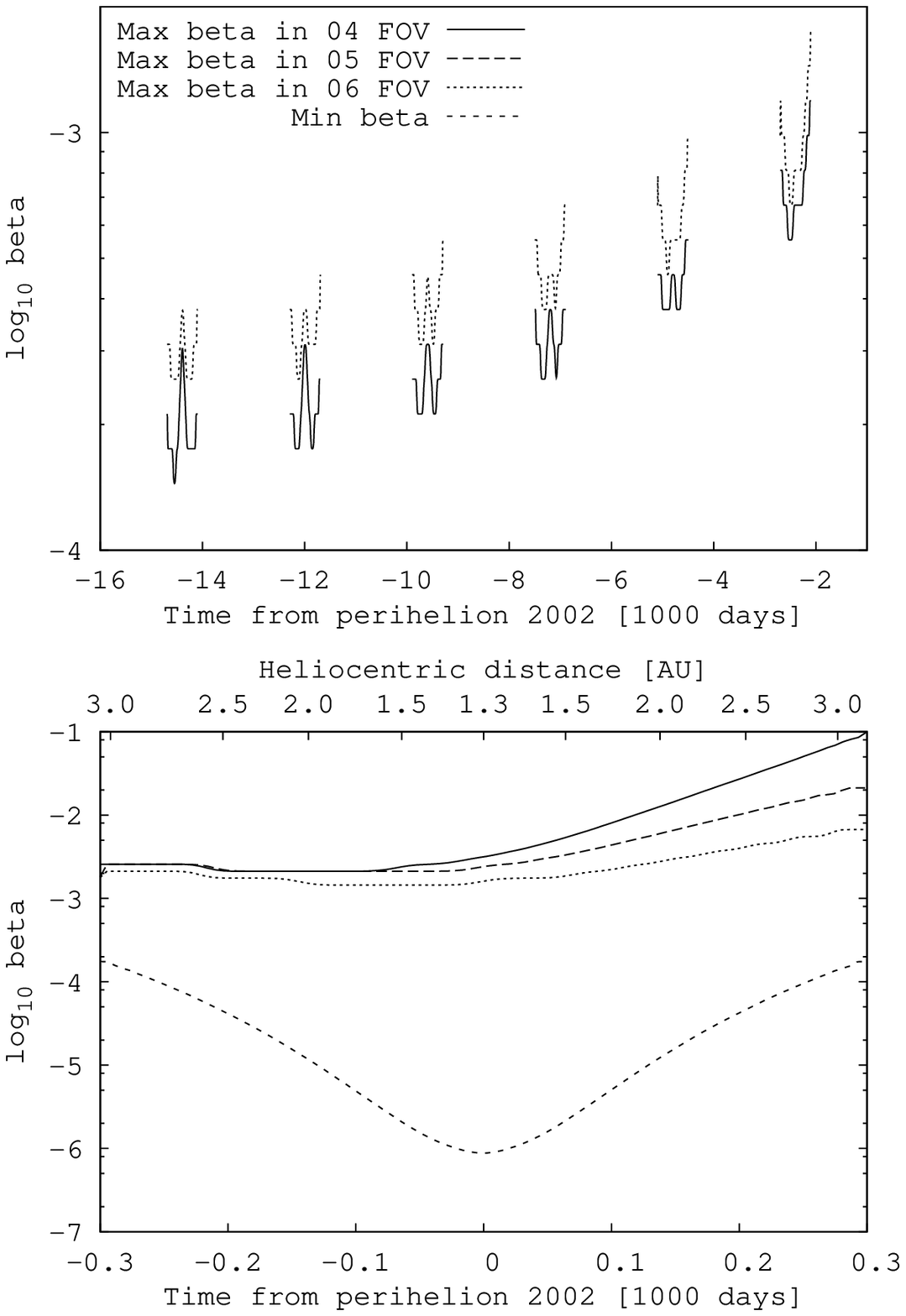}
\caption{Minimum and maximum radiation pressure coefficient of particles in the FOVs of our observations as functions of the emission time for $f_{\rm v}/\sqrt{Q_{\rm pr}}=1.6$. The upper panel shows the maximum $\beta$ of particles remaining in the FOVs of our observations for emission between 1959 and 1997. The lower panel shows the minimum and maximum $\beta$ of particles from the apparition immediately preceding our observations (2002/03).
The 2005 maximum curve (not plotted in the upper panel) is between the 2004 and 2006 curves, and the minimum curve is the same during all apparitions.} 
\label{fig:minmax_beta}
\end{figure}

Figure~\ref{fig:minmax_beta} shows the $\beta$-range of particles visible in the FOV of each observation as a function of the emission time. In general, the longer the emission lies back in time, the larger are the smallest particles remaining in the FOV, because of the action of radiation pressure. 
The minimum $\beta$ (largest particles) depends on the assumed escape speed from the nucleus and the adopted model of emission speeds (Eq.~\ref{eq:v_of_beta}). It is the same for all observations and apparitions. The maximum $\beta$ depends on the size of the FOV, on the observation time, and on the assumed $v(\beta,r_{\rm h})$.

\begin{figure}[]
\center
\includegraphics[clip,width=.5\textwidth]{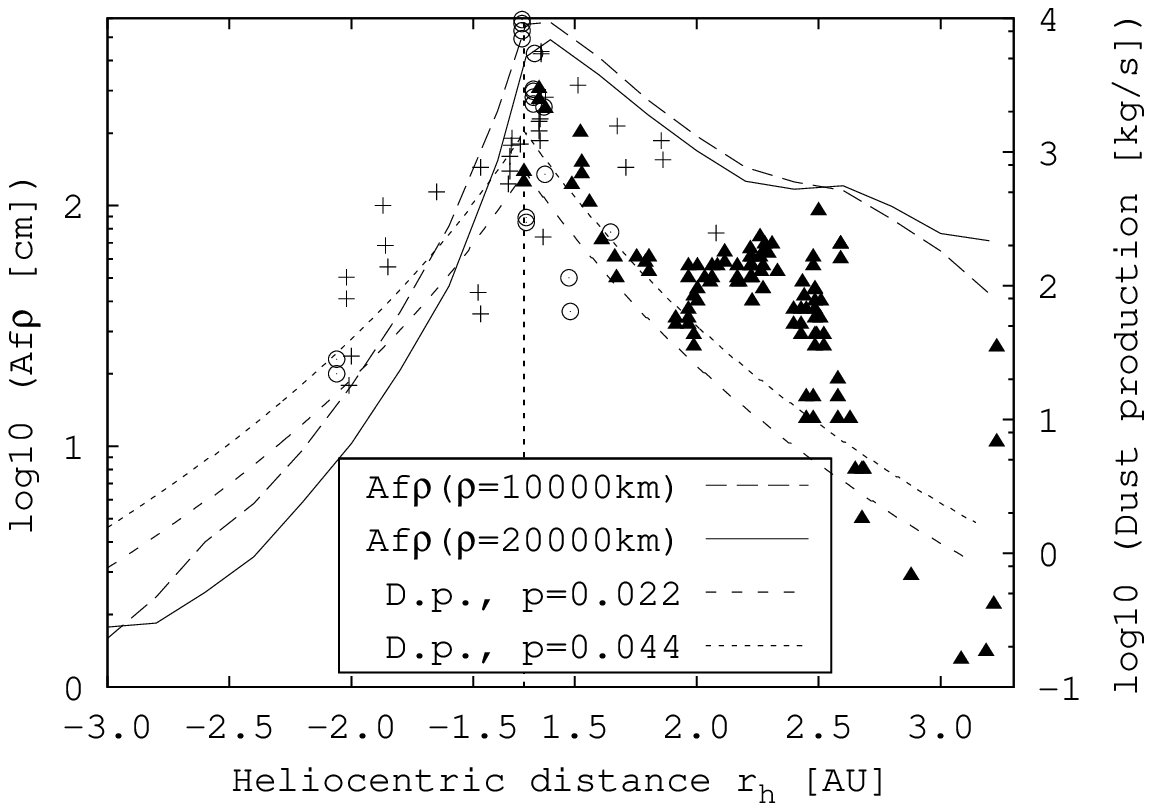}
\caption{Minimum \afr\ parameter and mass production rates for the best-fitting simulation parameters ($a=0.1$ and $f_{\rm v}/\sqrt{Q_{\rm pr}}=1.6$) and particles in the $\beta$ range of 2004 as shown in Fig.~\ref{fig:minmax_beta}. 
The \afr\ was measured at two nuclear distances ($\rho$ = 10000\,km and $\rho$ = 20000\,km) in simulated images at various heliocentric distances. The plotted values are assuming a constant phase function. For comparison we plot observed values of \afr: crosses are data from the 1982/83 apparition (\cite{storrs-cochran1992a, feldman-ahearn2004, schleicher2006}), open circles are data from 1996/97 (CARA data (\mbox{http://berlinadmin.dlr.de/Missions/corot/caesp/comet\_db.shtml}); \cite{weiler-rauer2004a,schleicher2006}), triangles are from 2002/03 (CARA; \cite{kidger2003, schulz-stuewe2004a, kelley-reach2008}).  
The mass production rate is plotted for the two extreme values of the albedo shown in Fig.~\ref{fig:albedo}.}
\label{fig:afr}
\end{figure}

Figure~\ref{fig:afr} shows the minimum \afr\
corresponding to our best-fitting model as a function of the heliocentric distance. To obtain the \afr\ plot, we have simulated images for a set of observation dates taking into account for emission only particles within the $\beta$-range depicted by the 2004 curve (solid) in Fig.~\ref{fig:minmax_beta}. We used the 2004 curve because it corresponds to the largest minimum \afr\ of the three. The actual \afr\ at a given time may have been higher if smaller particles were present that have left the FOV by the time of our observations. The \afr-curves shown in Fig.~\ref{fig:afr} do not reflect so much the activity of the comet as the dynamics of particles moving away from the comet. For comparison, we plot observed \afr\ data: our model predicts \afr\ on the same order of magnitude or larger than the observations, which we discuss in more detail in Section~\ref{sec:results}.
Figure~\ref{fig:afr} also shows the minimum dust mass production rate for geometric
albedos of 0.022 and 0.044 and $Q_{\rm pr} =1$. The depicted production rates correspond to dust in the same $\beta$-range as used for \afr. 

Figure~\ref{fig:beta_distance} shows the distribution of particle $\beta$ coefficients along the trail for simulations with the best-fitting parameters. Close to the nucleus, the trail consists of particles having approximately $10^{-6}<\beta<10^{-4}$. With increasing distance from the nucleus, the particles have higher $\beta$.  At a given nucleus distance, particles from the most recent apparition have higher $\beta$ than from previous apparitions: particles in the ``old trail'' have $\beta < 0.001$, while ``neckline'' particles are in the range of $0.001 \leq \beta \leq 0.01$ in the Spitzer images, and $0.001 \leq \beta \leq 0.1$ in the 2004 WFI image. The contribution of particles with $\beta > 0.01$ to the neckline brightness in 2004 is 10\%\ or less.

\section{Discussion}
\label{sec:discussion}


We have investigated the properties of large dust particles emitted by comet
67P/Churyumov-Gerasimenko through analysis of three images of its
dust trail -- one in visible light and two at 24\,\micron\ -- obtained at
heliocentric distances between 4.7\,AU and 
5.7\,AU. We have generated simulated images of the
cometary trail, and derived model parameters suitable to reproduce the observations.

Assuming that the comet is not active outside 3.1\,AU, the dust present in the
images is characterised by radiation pressure parameters $\beta <$ 0.01 in the 
infrared images, and $\beta <$ 0.1 in the visible image. For a bulk density of 1000\,kg/m$^3$, this corresponds to minimum sizes of 60\,\micron\ and 6\,\micron, respectively.
Our data do not allow us to make any conclusions about grains having higher $\beta$.

Our model has four free parameters: the relation between emission speeds and radiation pressure coefficient $\beta$, the exponent of the 
dust $\beta$ distribution (or -- for size-independent bulk density -- differential size distribution), the production rate of dust cross-section, and 
the average geometric albedo of the dust at visible wavelengths.

The closest reproduction of the observations was achieved with a $\beta$
distribution exponent of $a = 0.1$ (differential size distribution exponent $\alpha = -4.1$), and $f_{\rm v}/\sqrt{Q_{\rm pr}}=1.6$ in Eq.~\ref{eq:v_of_beta}. A particle having $\beta=0.01$ is emitted with 25\,m/s at perihelion and with 2\,m/s at 3\,AU.
The production rates of particles larger than 10\,\micron\ are on the order of
1000 - 2000 kg/s at perihelion and of 1 - 2 kg/s at 3\,AU.

\begin{figure}[]
\includegraphics[clip,width=0.5\textwidth]{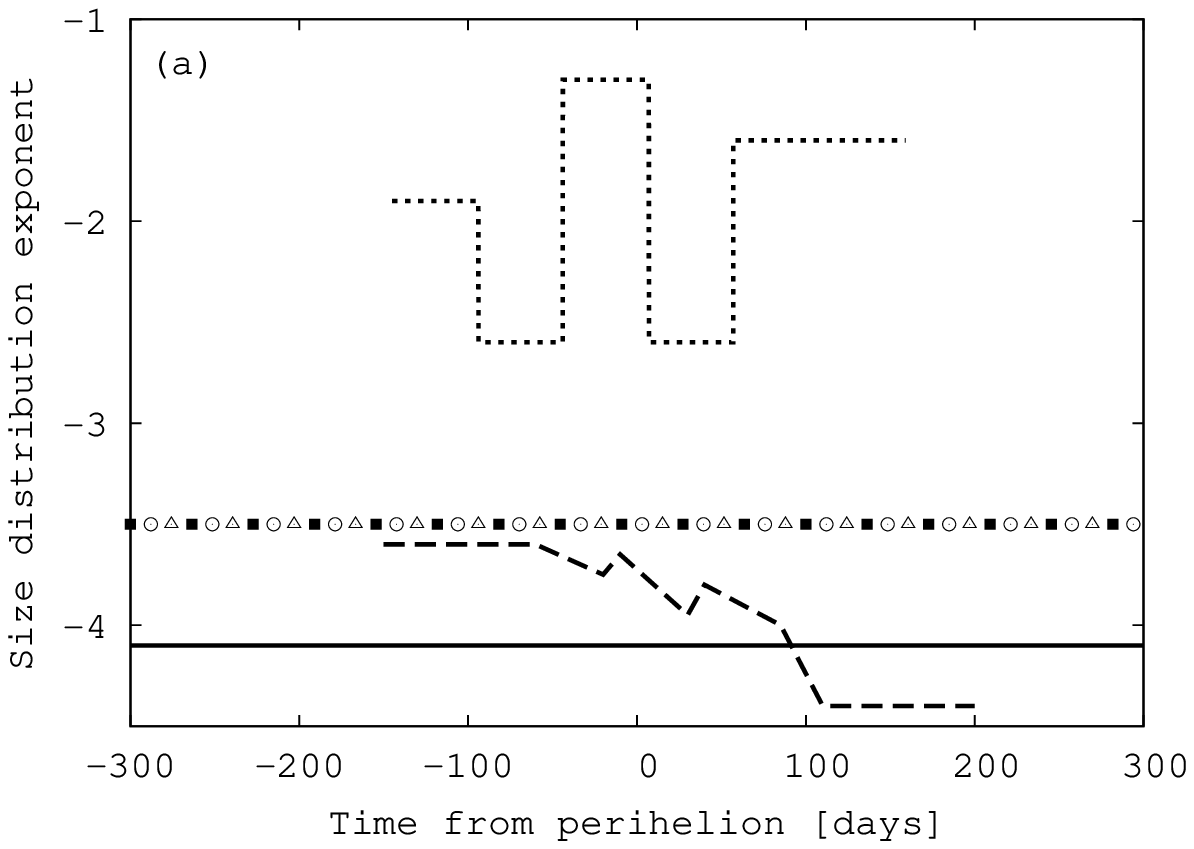}
\includegraphics[clip,width=0.5\textwidth]{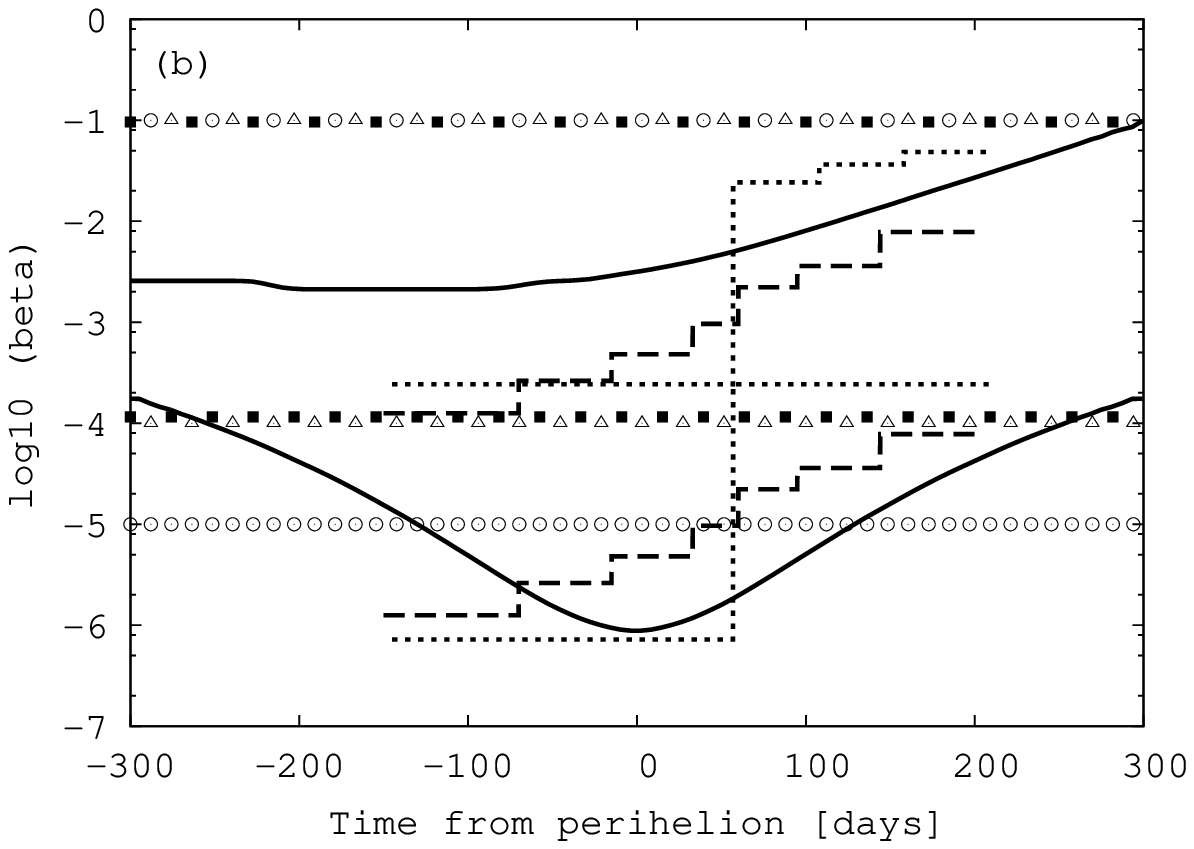}
\includegraphics[clip,width=0.5\textwidth]{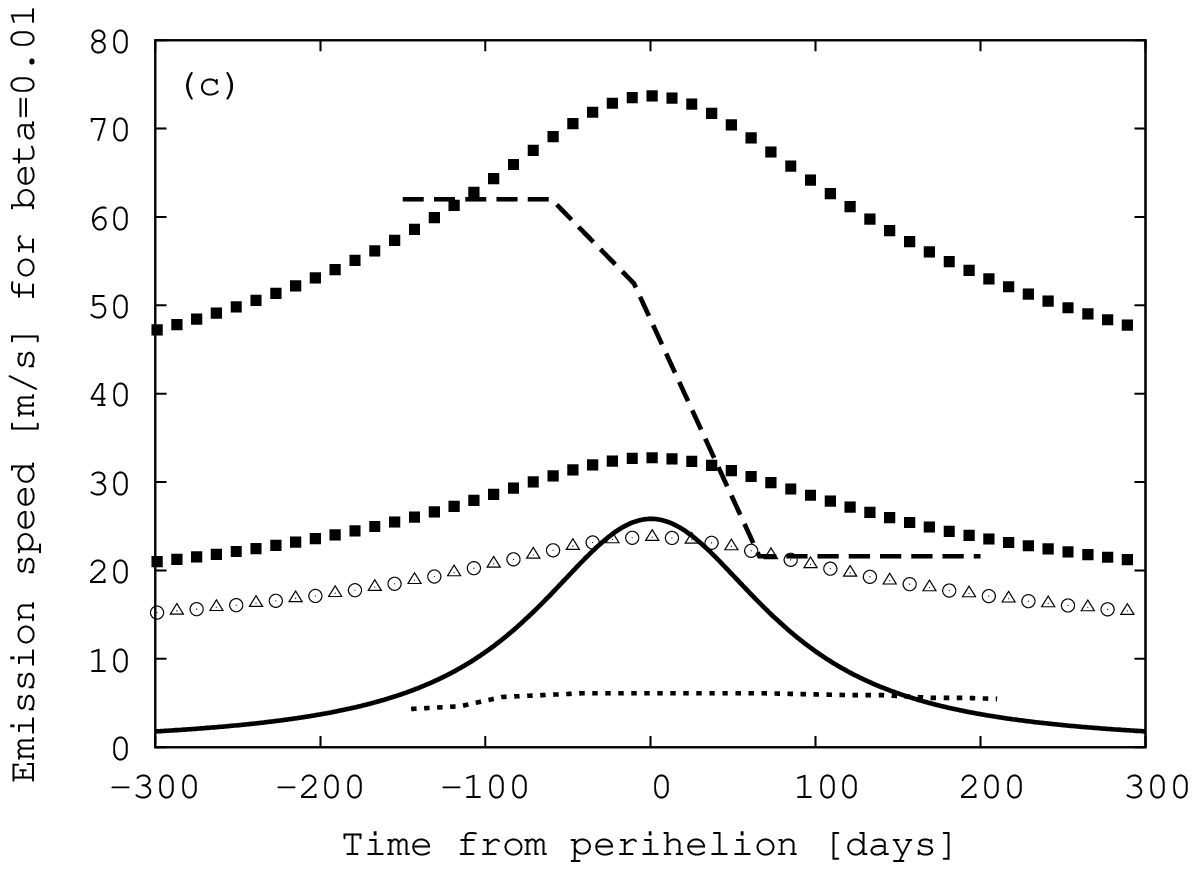}
\includegraphics[clip,width=0.5\textwidth]{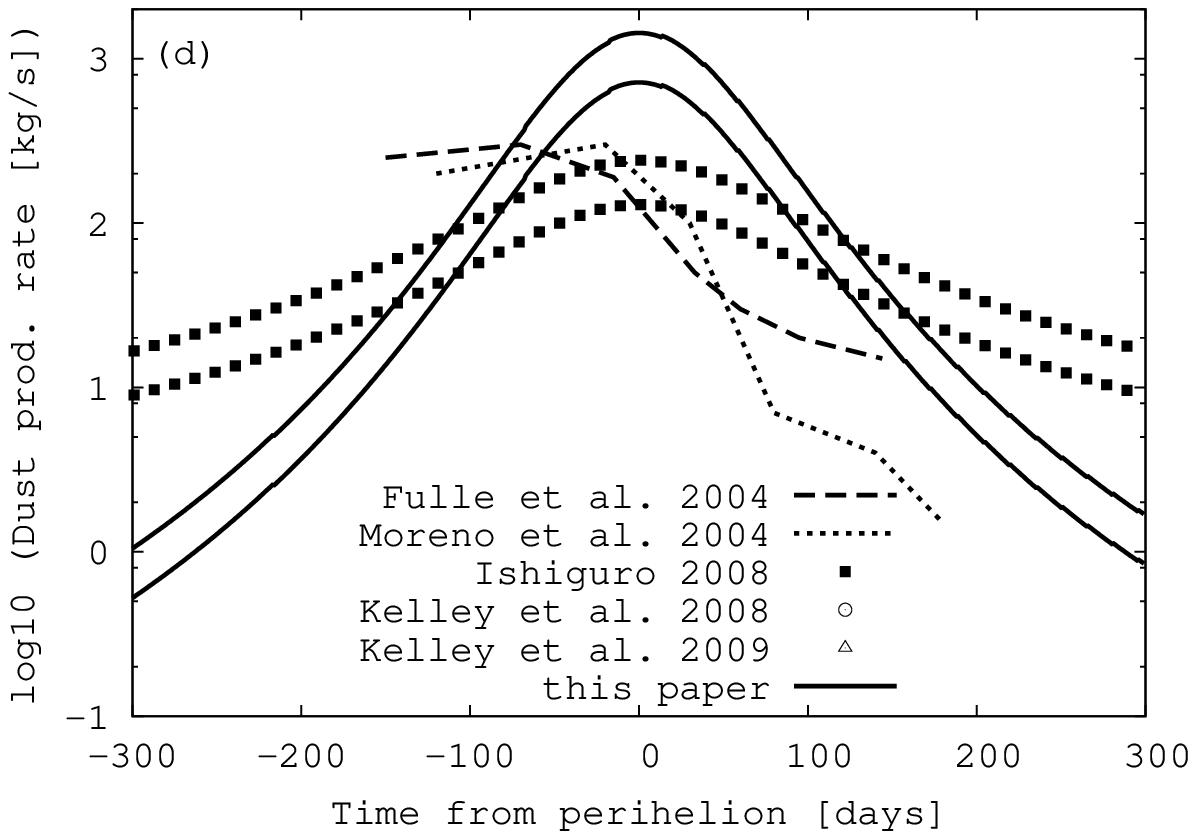}
\caption{Comparison of published models of the dust emission from comet \cg\ derived from tail or trail images. Where two or more models use the same value, the lines are plotted with a small offset to improve visibility. a) Exponent of the differential size distribution. b) Minimum and maximum $\beta$ as a function of time. For \cite{fulle-barbieri2004a, moreno-lara2004} and this paper, the $\beta$ range used to fit the images is plotted. \cite{ishiguro2008} and \cite{kelley-reach2008, kelley-wooden2009} do not give details on the time dependence of the $\beta$ range. c) Emission speeds. For non-isotropic models (\cite{moreno-lara2004, ishiguro2008, kelley-reach2008, kelley-wooden2009}) the speed at the subsolar point is given. d) Mass production rate in the $\beta$ range shown in panel b). \cite{ kelley-reach2008, kelley-wooden2009} use a production rate proportional to $r_{\rm h}^{-5.8}$ but do not mention absolute values.}
\label{fig:models_comparison}
\end{figure}

Our model reproduces the surface brightness of the trail seen with
Spitzer in February 2004 (\cite{kelley-reach2008}) and of the trail profiles obtained with VLT in April and June 2004 (\cite{tubianaPhD}) within a factor of two. It also reproduces the surface brightness of coma
and tail in March 2003 (\cite{fulle-barbieri2004a, moreno-lara2004}), if taking into account only particles in the $\beta$ range depicted in Fig~\ref{fig:minmax_beta}. This implies that our model does not allow for additional, smaller particles being emitted from the comet, which at that time was active.
%
We have simulated our images with the parameters derived by
\cite{fulle-barbieri2004a} (see Fig.~\ref{fig:models_comparison}) and found that their model underestimates the
surface brightness in our data by an order of magnitude.

The production rates, emission speeds, and $\beta$ distribution derived in Section~\ref{sec:results} correspond to \afr\ on the order of 300\,cm at perihelion and 50\,cm at 3\,AU post-perihelion. These values include only particles in the $\beta$ range shown in Fig.~\ref{fig:minmax_beta}, i.e. having $\beta<0.1$.
Thus, with our model, the production of large particles alone leads to \afr\  
on the same order of magnitude or larger than the measured \afr.
The $\beta$ distribution exponent of $a=0.1$ implies that in our model the largest contribution to \afr\ stems from the smallest particles present. Since the particles with $\beta<0.1$ more than reproduce the observed \afr, we conclude that the exponent $a=0.1$ cannot be valid for smaller particles. 
Particles with $\beta>0.01$ are in our data only present at distances larger that 5\arcmin\ behind the nucleus in the 2004 image. In our best-fitting model they contribute about 10\% to the flux from this region (Fig.~\ref{fig:beta_distance}), and the model slightly overestimates the total flux (Fig.~\ref{fig:best_fit_04}). 
If the emission of particles with $0.01<\beta<0.1$ was less than given by the assumed power-law, we expect that both the simulated \afr\ in Fig.~\ref{fig:afr} and the simulated trail in Fig.~\ref{fig:best_fit_04} would be closer to the observations, while the simulations of the Spitzer images would not change.
We conclude that the coma brightness while the comet was in the inner solar system may have been dominated by particles with $\beta \approx 0.01$ (60 - 600\,\micron, for densities between 1000 and 100\,kg/m$^3$). 

\cite{hanner-tedesco1985a} in thermal infrared spectra of the \cg\ coma have found excess colour temperatures that are consistent with the temperature excess observed in the trail by \cite{sykes-walker1992a}. All measurements can be described by Eq.~\ref{eq:T_eq} and $Z=0.6\pm0.2$ (\cite{agarwal-mueller2009}). 
\cite{divine-fechtig1986} explain the excess colour temperature in the coma with a dominant particle size of 0.6\,\micron. Alternative explanations could include grains sufficiently large to sustain a temperature gradient, or a certain degree of porosity.

A different explanation for the apparently high production rates of large particles required to reproduce the trail images, would be the fragmentation of large particles several years after their emission. If the 
trail particles had still been disintegrating at the time of our observations,
there would have been a population of small
particles in the FOVs that may have contributed significantly to the measured
surface brightness. Thus the amount of large particles originally produced from the comet could have been smaller, leaving room for a contribution of smaller particles to the coma flux.
The cause of a possible fragmentation remains at present speculative.

Although we have assumed isotropic dust emission from the comet,
our method primarily constrains the component of the emission velocity
perpendicular to the Sun-comet axis: The position of a grain relative to the
nucleus along the orbit depends mainly on its velocity component parallel to the orbital motion of the nucleus at the emission time, while its distance from the projected orbit depends on the velocity component perpendicular to the comet's orbital plane and on the time elapsed since emission. The dust emission speeds close to the subsolar point are therefore less constrained by our model. 
We expect that an insolation-driven emission model, where the local production rates and speeds depend on the solar zenith angle, will also reproduce the observations, as long as the velocity components perpendicular to the comet-Sun vector are similar to those we found with the isotropic model.
%
A different possible non-isotropic scenario could include high emission rates of particles with a large velocity component parallel to the direction of motion of the comet. Such particles would eventually fall back behind the nucleus and could form the observed bulge. 

To put our findings in the context of published models we show in 
Fig.~\ref{fig:models_comparison} a comparison of the results from different models for the dust size distribution, $\beta$ range, emission speed, and dust production. Details of the observations on which the different models were based are listed in Table~\ref{tab:observations_summary}.
The time-dependence of all parameters is a free parameter for \cite{fulle-barbieri2004a} and \cite{moreno-lara2004}, while \cite{ishiguro2008, kelley-reach2008, kelley-wooden2009} and our model assume variable but time-independent size distributions and power-laws with heliocentric distance with variable exponents for emission speed and dust production. \cite{ishiguro2008} and \cite{kelley-reach2008, kelley-wooden2009} find a size distribution exponent of -3.5, for \cite{fulle-barbieri2004a} it drops from -3.5 before perihelion to -4.4 after. Their intermediate values around perihelion are similar to our result of -4.1. In all models -- with the exception of \cite{moreno-lara2004} -- the dust cross-section is dominated by the smallest particles ($\alpha<-3$).
The $\beta$ range given by \cite{fulle-barbieri2004a, moreno-lara2004} and us reflects mainly the dates and FOVs of the modelled observations, with an additional influence of the emission speed (particles emitted at higher speed parallel to the orbital motion disappear more quickly from a given FOV). In our model, the largest grain size depends on the assumed escape speed of the comet and the adopted $v(\beta)$. \cite{ishiguro2008} and \cite{kelley-reach2008, kelley-wooden2009} seem to assume fixed size ranges: due to the shallow dependence of the emission speed on heliocentric distance in their models ($r_{\rm h}^{-0.5}$), the speed of a particle with $\beta = 10^{-4}$ is larger than the escape speed even at aphelion. 
The emission speeds in all models except that of \cite{moreno-lara2004} are roughly consistent at perihelion, but differ by about an order of magnitude at 3\,AU. \cite{ishiguro2008} and \cite{kelley-reach2008, kelley-wooden2009} assume $v \propto r_{\rm h}^{-0.5}$, which represents the varying insolation with $r_{\rm h}$. We use  $v \propto r_{\rm h}^{-3}$ based on the observed gas production rate of the comet. 
In the model of \cite{fulle-barbieri2004a} the time dependence of the emission speed is a free parameter, and they find a strong decrease from before to after perihelion, which would indicate that also the gas production should be asymmetric about perihelion. Note that $v \propto \beta^{1/2}$ in all models except for \cite{moreno-lara2004} who use a variable exponent and find fitting solutions also with $v \propto \beta^{1/6}$. 
Similar to the speeds, the dust production rates are free to vary with time in \cite{fulle-barbieri2004a} and \cite{moreno-lara2004}, who find a strong asymmetry with perihelion. \cite{ishiguro2008, kelley-reach2008, kelley-wooden2009} and our model assume a power-law dependence on heliocentric distance. \cite{ishiguro2008} treats the exponent as a variable parameter and finds -3 to fit his images best. \cite{kelley-reach2008, kelley-wooden2009} assume an exponent of -5.8, but do not give absolute production rates. Our model assumes an exponent of -8. The dust production per orbit in our model is about 2.5 times higher than found by \cite{ishiguro2008}.

We wish to stress that in this paper we have presented one possible
model that reproduces the discussed three observations and does not entirely
fail for a set of others. We have assumed fixed dependencies of emission speed
and production rate on the heliocentric distance, a constant $\beta$ distribution
exponent, isotropic emission, and size-independent optical properties. We have done this because the observational evidence to support more elaborate models is scarce. However, models with different assumptions on e.g. the dependencies on heliocentric distance and the emission geometry reproduce the respective observations equally successfully (\cite{fulle-barbieri2004a, moreno-lara2004, ishiguro2008, kelley-reach2008, kelley-wooden2009}). It must be the task for future work to find ways of discriminating between the differences in the existing models.
A key to such effort lies in the increasing amount of observations of comet \cg, characterising both the time-dependent activity of the nucleus and the temporal evolution of the dust trail.

\appendix

\section{Mathematical method}
\label{app:math}
We describe in this appendix the algorithm to simulate the image of a shell of particles characterised by the radiation pressure parameter $\beta$, the emission time $t_{\rm e}$, and the isotropic emission speed $v_{\rm e}$ for a given observation time and observer location. 
\input{math_method}
\clearpage
\input{math_solution-general}
\input{math_solution-hom-emission}

\ack
We wish to thank Ren{\'e} Laureijs (ESA/ESTEC) for his help to improve this paper. J.A. is grateful to Marco Fulle (INAF) for extensive discussions on the subject of dust modelling. We thank Masateru Ishiguro (Seoul National University) and one anonymous referee for their comments that helped to improve the manuscript.
This work is based on observations made with the MPG/ESO 2.2m telescope at the
La Silla Observatory under programme ID 072.A-9011(A), and 
with the Spitzer Space Telescope under programme ID 20235. Spitzer is
operated by the Jet Propulsion 
Laboratory, California Institute of Technology under a contract with
NASA. For this research we have made 
use of the USNOFS Image and Catalogue Archive operated by the United States
Naval Observatory, Flagstaff Station. 
Ephemerides were obtained from the HORIZONS system provided by the Solar
System Dynamics Group of the Jet Propulsion Laboratory. The data reduction was made with IRAF, which is written and supported by the IRAF
  programming group at the National Optical Astronomy Observatories (NOAO) in
  Tucson, Arizona, and with MOPEX, provided by the Spitzer Science Center.

\label{lastpage}


\bibliography{/usr4/users/jagarwal/Latex/refs}

\bibliographystyle{plainnat}

\end{document}

%% file: results_04.tex
\begin{table}[t]
\caption{Simulation results for the WFI observation in April 2004. We have run simulations for 21 values of the parameter $f_v/\sqrt{Q_{\rm pr}}$ and five values of the $\beta$-distribution exponent $a$. 
For each pair of parameters we have plotted profiles like Fig.~\ref{fig:best_fit_04} and by visual inspection compared the simulation and observation with respect to five criteria: ``P'', ``F'', and ``D'' were evaluated in a profile like in panel (a), and ``T'' and ``N'' refer to the trail width as in panel (b).
{\bf P} is the position of the peak brightness near the nucleus. ``-'' (``+'') means that the peak is too far behind (in front of) the nucleus in the direction of its orbital motion.
{\bf F} characterises the brightness in front of the nucleus (positive x-values). ``+'' (``-'') indicate that the trail in this region is too bright (faint). 
{\bf D} evaluates the trail brightness in the nearly flat part of the profile far behind the nucleus. ``+'' (``-'') have the same meaning as for ``F''.
{\bf T} and {\bf N} refer to the width on the trail (``+'') and neckline (``x'') side of panel (b). ``+'' (``-'') means that the simulation is wider (narrower) than the observation. In all cases, ``O'' means that the simulation reproduces the observation within the errorbars. ``n.a.'' for P indicates that the profile is too flat to identify a peak. Good fits are shadowed in grey. 
}
\vspace{2mm}
\label{tab:profile_eval_04}
\begin{tiny}  
\begin{tabularx}{\textwidth}{l|*{5}{X}|*{5}{X}|*{5}{X}|*{5}{X}|*{5}{X}}
\hline\noalign{\smallskip}
$(f_v/\sqrt{Q_{\rm pr}}) \: \backslash a$ &\multicolumn{5}{|c|}{-0.4}& \multicolumn{5}{|c|}{-0.2} &\multicolumn{5}{|c|}{0}  & \multicolumn{5}{|c|}{0.2} &\multicolumn{5}{|c}{0.4}\\
\hline\noalign{\smallskip}
&P&F&D&T&N&P&F&D&T&N&P&F&D&T&N&P&F&D&T&N&P&F&D&T&N\\
\hline\hline\noalign{\smallskip}
0.2 & O &  -- &  O &  -- &   O 
& O &  -- &   + &  -- &   O 
&  -- &  -- &   + &  -- &   O 
&  -- &  -- &   + &  -- &   O 
&  -- &  -- &   + &  -- &   O \\ 
0.4 &   O & O &  O &   O &  -- 
& O & O &  O &  -- &   O 
& O & O &   + & O &   O 
& O & O &   + &  -- &   O 
&  -- &  -- &   + &  -- &   O \\ 
0.6 &   O & O &   O &   O &  -- 
&\gr   O &\gr  O &\gr   O &\gr    O &\gr    O 
& O & O &   + &   O &   O 
& O & O &   + & O &   O 
&  -- &  -- &   + &  -- &   O \\ 
0.8 &   O &   O & O &   O &  -- 
&\gr    O &\gr  O & \gr   O & \gr   O & \gr   O 
&   O & O &   + &   O &   O 
& O & O &   + &   O &   O 
&  -- & O &   + & O &   O \\ 
1.0 -- 1.2&   O &   O & O &   O &  -- 
& \gr   O &\gr    O &\gr    O &\gr    O &\gr  O 
&   O &   O &   + &   O &   O 
&   O & O &   + &   O &   O 
& O & O &   + &   O &   O \\ 
1.4 -- 1.8 &   O &  O & O &   O &  -- 
&\gr O &\gr     O &\gr    O &\gr    O &\gr  O 
&\gr    O &\gr    O &\gr   O &\gr    O & \gr   O 
&   O &   O &   + &   O &   O 
& O & O &   + &   O &   O \\ 
2.0 &  O &  O &  -- &   O &  -- 
&\gr   O &\gr   O &\gr    O & \gr   O & \gr   O 
&\gr    O &\gr   O &\gr   O &\gr    O & \gr   O 
&   O &  O &   + &   O &   O 
&   O &   O &   + &  O &   O \\ 
2.2 &   + &   + & O &   O &  -- 
& \gr  O &\gr   O &\gr    O &\gr    O &\gr  O 
&\gr   O &\gr   O &\gr   O &\gr    O & \gr   O 
&   O &  O &   + &   O &   O 
&   O &  O &   + &   O &   O \\ 
2.4 &   + &   + & O &   O &  -- 
&  O &   + &   O &   O & O 
&\gr   O & \gr  O & \gr  O &\gr    O & \gr   O 
&   O &  O &   + &   O &   O 
&   O &  O &   + &   O &   O \\ 
2.6 &   + &   + & O &   O &  -- 
&  O &   + &   O &   O &   O 
&  O &   + &  O &   O &   O 
&   O &  O &   + &   O &   O 
&   O &  O &   + &  O &   O \\ 
2.8 &   + &   + &  -- &   O &  -- 
&   + &   + &   O &   O &   O 
&  O &   + &  O &   O &   O 
&  O &   + &   + &   O &   O 
&   O &   + &   + &  O &   O \\ 
3.0 - 4.0 &   + &   + &  -- &   O &  -- 
&   + &   + &   O &   O &   O 
&   + &   + &  O &   O &   O 
&  O &   + &   + &  O &   O 
&  O &   + &   + &  O &   O \\ 
5.0 &   + &   + &  -- &   O &  -- 
&   + &   + &   O &   O &   O 
&   + &   + &  O &  O &   O 
&  na &   + &   + &  O &   O 
&  na &   + &   + &  O &   O \\ 
6.0 &   + &   + &  -- &   O &  -- 
&   + &   + &   O &  O &   O 
&  na &   + &  O &  O &   O 
&  na &   + &   + &  O &   O 
&  na &   + &   + &  O &   O \\ 
8.0 -- 10.0 &   + &   + &  -- &   O &  -- 
&   + &   + &   O &  O &   O 
&  na &   + &  O &  O &   O 
&  na &   + &   + &  O &   O 
&  na &   + &   + &   + &   + \\ 
\noalign{\smallskip}\hline
\end{tabularx}
\end{tiny}
\end{table}

%% file: results_05.tex
\begin{table}[t]
\caption{Simulation results for the Spitzer observation of August 2005. For details see Table~\ref{tab:profile_eval_04}.}
\vspace{2mm}
\label{tab:profile_eval_05}
\begin{tiny}  
\begin{tabularx}{\textwidth}{l|*{5}{X}|*{5}{X}|*{5}{X}|*{5}{X}|*{5}{X}}
\hline\noalign{\smallskip}
$(f_v/\sqrt{Q_{\rm pr}}) \: \backslash a$ &\multicolumn{5}{|c|}{-0.4}& \multicolumn{5}{|c|}{-0.2} &\multicolumn{5}{|c|}{0}  & \multicolumn{5}{|c|}{0.2} &\multicolumn{5}{|c}{0.4}\\
\hline\noalign{\smallskip}
&P&F&D&T&N&P&F&D&T&N&P&F&D&T&N&P&F&D&T&N&P&F&D&T&N\\
\hline\hline\noalign{\smallskip}
0.2 &   O &  -- &   O &  -- & O 
& O &  -- &   O &  -- & O 
&  -- &  -- &  -- & O &     
&  -- &  -- &   + &  -- & O 
&  -- &  -- &   + &  -- & O \\ 
0.4 &   O &  -- & O &  -- &  -- 
&   O &  -- & O &  -- & O 
&   O &  -- &   O &  -- & O 
& O &  -- &   + &  -- & O 
&  -- &  -- &   + &  -- & O \\ 
0.6 &  O & O &  -- &  -- &  -- 
&  O & O & O &  -- &  -- 
&   O &  -- & O &  -- & O 
&   O &  -- &  O &  -- & O 
&  -- &  -- &   + &  -- & O \\ 
0.8 &  O & O &  -- &  -- &  -- 
&  O & O & O &  -- &  -- 
&  O & O & O &  -- & O 
&   O & O &   O &  -- & O 
& O &  -- &   + &  -- & O \\ 
  1.0 &   + &   O &  -- &  -- &  -- 
&  O &   O & O &  -- &  -- 
&  O &   O & O &  -- & O 
&   O & O &   O &  -- & O 
&   O & O &   + &  -- &   O \\ 
1.2 -- 1.4 &   + &   O &  -- &  -- &  -- 
&   + &   O &  -- &  -- &  -- 
&  O &   O & O &  -- & O 
&   O &   O &   O &  -- & O 
&   O & O &   + &  -- & O \\ 
1.6 &   + &  O &  -- & O &  -- 
&   + &  O &  -- & O &  -- 
&   + &  O & O & O & O 
&\gr O &\gr  O &\gr   O &\gr O &\gr O 
&   O &   O &   + & O & O \\ 
1.8 &   + &   + &  -- & O &  -- 
&   + &   + &  -- & O &  -- 
&   + &   + & O & O & O 
&  O &   + &   O & O & O 
&   O &  O &   + & O &   O \\ 
2.0 -- 2.8 &   + &   + &  -- & O &  -- 
&   + &   + &  -- & O &  -- 
&   + &   + & O &   O & O 
&   + &   + &   O &   O &   O 
&   O &   + &   + &   O &   O \\ 
3.0 &   + &   + &  -- &   O &  -- 
&   + &   + &  -- &   O & O 
&   + &   + & O &  O &   O 
&   + &   + &   O &  O &   O 
&  O &   + &   + &   + &  O \\ 
3.5 &   + &   + &  -- &   O &  -- 
&   + &   + &  -- &   O & O 
&   + &   + & O &  O &   O 
&   + &   + &   O &   + &  O 
&  O &   + &   + &   + &  O \\ 
  4.0 &   + &   + &  -- &   O &  -- 
&   + &   + &  -- &  O & O 
&   + &   + & O &   + &   O 
&   + &   + &   O &   + &   + 
&  O &   + &   + &   + &   + \\ 
  5.0 &   + &   + &  -- &   O &  -- 
&   + &   + & O &  O &   O 
&   + &   + &   O &   + &   + 
&   + &   + &  O &   + &   + 
&  O &   + &   + &   + &   + \\ 
  6.0 &   + &   + &  -- &   O &  -- 
&   + &   + & O &   + &  O 
&   + &   + &   O &   + &   + 
&   + &   + &   + &   + &   + 
&  na &   + &   + &   + &   + \\ 
  8.0 &   + &   + &  -- &   O & O 
&   + &   + &  -- &   + &   + 
&   + &   + &   O &   + &   + 
&  na &   + &   + &   + &   + 
&  na &   + &   + &   + &   + \\ 
 10.0 &   + &   + &  -- &   O & O 
&   + &   + & O &   + &   + 
&   + &   + &   O &   + &   + 
&  na &   + &   + &   + &   + 
&  na &   + &   + &   + &   + \\ 
\noalign{\smallskip}\hline
\end{tabularx}
\end{tiny}
\end{table}

%% file: results_06.tex
\begin{table}[t]
\caption{Simulation results for the Spitzer observation of April 2006. For details see Table~\ref{tab:profile_eval_04}.}
\vspace{2mm}
\label{tab:profile_eval_06}
\begin{tiny}  
\begin{tabularx}{\textwidth}{l|*{5}{X}|*{5}{X}|*{5}{X}|*{5}{X}|*{5}{X}}
\hline\noalign{\smallskip}
$(f_v/\sqrt{Q_{\rm pr}}) \: \backslash a$ &\multicolumn{5}{|c|}{-0.4}& \multicolumn{5}{|c|}{-0.2} &\multicolumn{5}{|c|}{0}  & \multicolumn{5}{|c|}{0.2} &\multicolumn{5}{|c}{0.4}\\
\hline\noalign{\smallskip}
&P&F&D&T&N&P&F&D&T&N&P&F&D&T&N&P&F&D&T&N&P&F&D&T&N\\
\hline\hline\noalign{\smallskip}
0.2 &  -- &  -- &  O &  -- &  -- 
&  -- &  -- &  O &  -- &  -- 
&  -- &  -- &  O &  -- & O 
&  -- &  -- &   + &  -- & O 
&  -- &  -- &   + &  -- & O \\ 
0.4 &   O &  -- & O &  -- &  -- 
&   O &  -- &   O &  -- &  -- 
& O &  -- &  O &  -- & O 
&  -- &  -- &   + &  -- & O 
&  -- &  -- &   + &  -- & O \\ 
0.6 &   + &  -- &  -- &  -- &  -- 
&   O &  -- &   O &  -- &  -- 
&   O &  -- &   O &  -- & O 
& O &  -- &  O &  -- & O 
&  -- &  -- &   + &  -- & O \\ 
0.8 &   + & O &  -- &  -- &  -- 
&  O & O & O &  -- &  -- 
&   O & O &   O &  -- & O 
& O &  -- &   + &  -- & O 
&  -- &  -- &   + &  -- & O \\ 
  1.0 &   + & O &  -- &  -- &  -- 
&   + & O & O &  -- &  -- 
&   O & O &   O &  -- &  -- 
&   O & O &  O &  -- & O 
& O &  -- &   + &  -- & O \\ 
1.2 &   + &   O &  -- &  -- &  -- 
&   + &   O & O &  -- &  -- 
&  O & O &   O &  -- &  -- 
&   O & O &   + &  -- & O 
& O & O &   + &  -- & O \\ 
1.4 &   + &   O &  -- &  -- &  -- 
&   + &   O &  -- &  -- &  -- 
&  O &   O &   O &  -- &  -- 
&   O &   O &  O &  -- & O 
&   O & O &   + &  -- & O \\ 
1.6 &   + &   O &  -- & O &  -- 
&   + &   O &  -- & O &  -- 
&  O &   O &   O &  -- &  -- 
&\gr   O &\gr   O &\gr  O &\gr O &\gr O 
& O &   O &   + & O & O \\ 
1.8 &   + &  O &  -- & O &  -- 
&   + &  O &  -- & O &  -- 
&   + &  O & O & O &  -- 
&\gr   O & \gr  O & \gr O &\gr O & \gr O 
&   O &   O &   + & O & O \\ 
  2.0 &   + &  O &  -- & O &  -- 
&   + &  O &  -- & O &  -- 
&   + &  O &   O &   O & O 
&   + &  O &  O & O &   O 
&   O &  O &   + &   O &   O \\ 
2.2 &   + &  O &  -- & O &  -- 
&   + &   + &  -- & O &  -- 
&   + &   + & O & O & O 
& + &   O &  O &   O & O 
&   O &  O &   + &   O &   O \\ 
2.4 -- 2.8 &   + &   + &  -- & O &  -- 
&   + &   + &  -- &   O &  -- 
&   + &   + & O &   O & O 
&   + &   + &  O &   O & O 
&   O &   + &   + &   O &   O \\ 
  3.0 &   + &   + &  -- &   O &  -- 
&   + &   + &  -- &   O & O 
&   + &   + & O &   O &   O 
&   + &   + &   O &  O &   O 
&  na &   + &   + &  O &  O \\ 
3.5 &   + &   + &  -- &   O &  -- 
&   + &   + &  -- &   O &   O 
&   + &   + &   O &  O &   O 
&   + &   + &   O &  O &   O 
&  na &   + &   + &   + &  O \\ 
  4.0 &   + &   + &  -- &   O & O 
&   + &   + &  -- &  O &   O 
&   + &   + & O &  O &  O 
&   + &   + &   O &   + &  O 
&  na &   + &   + &   + &   + \\ 
  5.0 &   + &   + &  -- &   O & O 
&   + &   + &  -- &  O &  O 
&   + &   + & O &   + &   + 
&  na &   + &   O &   + &   + 
&  na &   + &   + &   + &   + \\ 
  6.0 &   + &   + &  -- &  O &   O 
&   + &   + & O &   + &  O 
&   + &   + &   O &   + &   + 
&  na &   + &   + &   + &   + 
&  na &   + &   + &   + &   + \\ 
  8.0 -- 10.0 &   + &   + &  -- &  O &   O 
&   + &   + & O &   + &   + 
&   + &   + &   O &   + &   + 
&  na &   + &   + &   + &   + 
&  na &   + &   + &   + &   + \\ 
\noalign{\smallskip}\hline
\end{tabularx}
\end{tiny}
\end{table}

%% file: math_method.tex
\subsection{Coordinate Systems}
\label{subsec:math_method_coords}
The coordinate system used to describe the image has its origin
at the comet nucleus, and $(x,y,z)$ form a right-handed, orthogonal system. The $z$-axis is along the line of sight, pointing away
from the observer. 
The $x$-axis is parallel to the Earth's equatorial plane, and $x$ increases with right
ascension (RA), from right (west) to left (east) for an observer looking at
the sky. 
The $y$-axis completes the orthogonal system, pointing northward in projection.
It is assumed that the $x$-$y$-plane corresponds to the
image plane, which introduces a distortion for larger fields of
view.

A second comet-centred coordinate system is the {\em Cometocentric Bipolar
System} (CBS) \citep{massonnePhD}. It is spanned by the orthogonal vectors
($\xi$,$\chi$,$\zeta$), 
where the $\zeta$-axis is along the
line Sun-comet, pointing away  from the Sun. The $\xi$-axis is in the orbital
plane of the comet, perpendicular to $\zeta$, and pointing in the direction of motion. $\chi$ completes the right-handed system.

\subsection{Reference Trajectories}
\label{subsec:math_method_reftraj}
A set of reference particles is defined that have the same $\beta$ and
$\te$ as the shell to be depicted, but not the same emission speed.
All reference particles have been emitted in the
direction of the heliocentric velocity vector of the comet ${{\mathbf e}_{\rm v,nucl}} (\te)$, with
varying absolute speeds.
One reference particle $J$ is needed for each pixel-column $x_j$ in the
image. 
To find the corresponding emission speeds $v_{\parallel,j}(x_j)$,
an auxiliary set of particles $i$ with
emission speeds $v_{\parallel,i}$ is defined and the particle positions $x_i$ in the image calculated.
The speeds of the reference particles $v_{\parallel,j}(x_j)$ are found by
interpolation of the relation $x_i (v_{\parallel,i})$.
The orbital elements of each reference particle $j$ are calculated from its  
heliocentric state vector at emission and $\beta$.
The coordinates ($x_j, y_j, z_j$) of the reference particles at the time of
observation are denoted by the vectors ${\mathbf r}_{0,j}$.

\subsection{Linearisation}
\label{subsec:math_method_lin}
The position ${\mathbf r} (\tobs)$ of a particle with emission
velocity ${\mathbf v}_{\mathrm e} = v_{\parallel,j} {{\mathbf e}_{\rm v,nucl}}
+ \Delta {\mathbf v}$ is 
linearised around ${\mathbf r}_{0,j}$ for not too large $\Delta{\mathbf v}$: 
\begin{equation}
\label{eq:lineq}
{\mathbf r} (\tobs) = {\mathbf r}_{0,j} (\tobs) + \Psij \Delta{\mathbf v}.
\end{equation}
The transformation matrix $\Psij$ depends on the trajectory of the
reference particle and on the dates of emission and observation. 
$\Psij$ is defined such that $\Delta{\mathbf v}$ is given in the
CBS-frame, whereas the positions ${\mathbf r}$ are in image coordinates.

If the particle trajectories are considered as Keplerian orbits with the
gravitational potential modified by the factor $\Delta\mu$ = $(1-\beta)$, the elements of the matrix
are obtained in analogy to \citet{massonnePhD}. 
While \citet{massonnePhD} uses the nucleus 
as reference object, we use dust particles. The full transformation formula is as 
follows \citep[pp. 127-128]{massonnePhD}: 
\begin{equation}
\label{eq:massonne_transf}
{\mathbf r}^{\rm \,CBS}(t) = \underbrace{
\frac{1}{\delta \tau / \delta t |_{\te}} 
\left(\!\! \begin{array}{ccc} 
\Phi_{22} & 0 & \Phi_{21} \\
0 & \Phi_{33} & 0 \\
\Phi_{12} & 0 & \Phi_{11}
\end{array} \!\!\right) {\mathbf v}_{\rm e}^{\rm \,CBS} (\te)}_{\rm I} 
\,+\, \underbrace{
a (1\!-\!e^2) \frac{\Delta \mu}{\mu} \left(\!\!\! \begin{array}{c}
\psi_2 \\ 0 \\ \psi_1 \end{array} \!\!\!\right)}_{\rm II}.
\end{equation}
Part II describes the change in trajectory due to the different strength of
radiation pressure for the nucleus and a dust particle. 
In the present approach, $\beta$ is the same for the reference particle
and the shell particles. Hence part II is always zero. 
In part I, $\delta \tau / \delta t |_{\te}$ is the time derivative of the
true anomaly of the reference object at the time of emission:
\begin{equation}
\delta \tau / \delta t |_{\te} = 
\sqrt{ \frac{\mu\, (1\!-\!\beta)}{a^3 (1\!-\!e^2)^3}} \:
(1 + e \cos \tau_{\rm e} )^2.
\end{equation}
$\mu$ = $G\Msun$ = 1.33 $\!\times\!$ 10$^{20}$ m$^3$/s$^2$ is the gravitational potential of the
Sun, and $a$, $e$, and $\tau$ are the semi-major axis, eccentricity and true
anomaly of the reference object, respectively. $\beta$ is the radiation
pressure coefficient of the reference object (i.\,e. $\beta$ = 0 in the situation
described by Massonne). 
The matrix elements $\Phi_{mn}$ are functions of $\tobs$,
$\te$, $\tau_{\rm obs}$, $\tau_{\rm e}$, $a$, and $e$. ${\mathbf r}^{\rm
  \,CBS}(t)$ and ${\mathbf v}_{\rm e}^{\rm \,CBS} (\te)$ are in CBS-coordinates
centred on the reference object at the times $t$ and $\te$, respectively.

The $\Delta{\mathbf v}$ of Equation~\ref{eq:lineq} is equivalent to 
${\mathbf v}_{\rm e}^{\rm \,CBS} (\te)$ of Equation~\ref{eq:massonne_transf}. It
follows that
\begin{equation}
\label{eq:psi_def}
\Psij = 
\frac{1}{\delta \tau_j / \delta t |_{\te}} \:
\Tj^{\,\rm\bf cbs-im}(\tobs) \:
\left(\!\! \begin{array}{ccc} 
\Phi_{22} & 0 & \Phi_{21} \\
0 & \Phi_{33} & 0 \\
\Phi_{12} & 0 & \Phi_{11}
\end{array} \!\!\right),
\end{equation}
where $\Tj^{\,\rm\bf cbs-im}(\tobs)$ is the matrix to
transform reference-particle centred CBS-coordinates to image coordinates
relative to the reference particle. The orbital elements in $\Psij$ and $\delta \tau_j / \delta t |_{\te}$ are those of the reference
particle, $\tau_j$, $a_j$, and $e_j$.

\subsection{Reference Frame in Ejection Velocity Space}
\label{subsec:math_method_refframe}
A reference frame in the space of ejection velocities is defined with unit vectors ${\mathbf u}_j$, ${\mathbf v}_j$, ${\mathbf w}_j$ such that 
\begin{eqnarray}
\label{eq:uvw_def_1}
\lambda \, {\mathbf e}_z &=& \Psij \, {\mathbf u}_j \\
\label{eq:uvw_def_2}
\nu \, {\mathbf e}_y + \mu \, {\mathbf e}_z &=& \Psij \, {\mathbf v}_j \\
\label{eq:uvw_def_3}
|{\mathbf u}_j| = |{\mathbf v}_j| &=& 1\\
\label{eq:uvw_def_4}
{\mathbf u}_j {\mathbf v}_j &=& 0\\
\label{eq:uvw_def_5}
{\mathbf w}_j &=& {\mathbf u}_j \times {\mathbf v}_j.
\end{eqnarray} 
A particle emitted with $\Delta{\mathbf v} = {\mathbf u}_j$ will be located on the same line of
sight as the reference particle, i.\,e. $x$ = $x_{0,j}$ and $y$ = $y_{0,j}$. For 
$\Delta{\mathbf v}$ = ${\mathbf v}_j$, the particle will have $x$ = $x_{0,j}$, but $y\neq
y_{0,j}$. To be consistent with the definition of $\Psij$, the unit vectors
${\mathbf u}_j$, ${\mathbf v}_j$, ${\mathbf w}_j$ need to be given in the CBS-frame at the
time $\te$. By $\Delta{\mathbf v}$ = $\rho {\mathbf u}_j + \sigma {\mathbf v}_j$, a plane in
the $\Delta{\mathbf v}$-space is defined that contains all possible $\Delta{\mathbf v}$ for
particles to lie in the image column characterised by $x$ = $x_{0,j}$.
If 
\begin{equation}
\Tj^{\,\rm\bf uvw-cbs}(\te) = \left( 
\begin{array}{ccc}
\label{eq:matrix_uvw2cbs}
u_{j,1}^{\rm cbs} & v_{j,1}^{\rm cbs} & w_{j,1}^{\rm cbs}\\
u_{j,2}^{\rm cbs} & v_{j,2}^{\rm cbs} & w_{j,2}^{\rm cbs}\\
u_{j,3}^{\rm cbs} & v_{j,3}^{\rm cbs} & w_{j,3}^{\rm cbs}
\end{array}\right)
\end{equation}
 is the matrix to
transform coordinates from the $(u,\!v,\!w)$-system to the CBS frame, the
matrix $\tilPsij$ = $\Psij \Tj^{\,\rm\bf   uvw-cbs}$ by definition takes the form
\begin{equation}
\tilPsij = 
\Psij \Tj^{\,\rm\bf uvw-cbs}=
\left(\begin{array}{ccc}
0 & 0 & c_1 \\
0 & \nu & c_2 \\
\lambda & \mu & c_3
\end{array}\right),
\end{equation}
where the vector ${\mathbf c}$ = $\Psij {\mathbf w}_j$ has been
defined. ${\mathbf c}$, $\lambda$, $\mu$, and $\nu$ depend on $j$, but
the index is omitted here for the sake of legibility.

\subsection{Mapping Emission Direction to Position in Image}
\label{subsec:math_method_projection}
Any vector ${\mathbf v}$ can be expressed as a sum of the unit vectors
${\mathbf u}_j$, ${\mathbf v}_j$, ${\mathbf w}_j$:
\begin{equation}
{\mathbf v} = v \sin \theta \sin \phi \, {\mathbf u}_j + 
v \sin \theta \cos \phi \, {\mathbf v}_j + v \cos \theta \, {\mathbf w}_j, 
\end{equation}
where $\theta \in [0,\pi]$ and $\phi \in [0,2\pi]$.
Hence, $\Delta {\mathbf v}$ is
\begin{eqnarray}
\nonumber \Delta {\mathbf v} &=& {\mathbf v}_{\rm e} -  v_{\parallel,j} \,{{\mathbf e}_{\rm v,nucl}} \\
&=& \ve \sin \theta \sin \phi \, {\mathbf u}_j + 
\ve \sin \theta \cos \phi \, {\mathbf v}_j + \ve \cos \theta \, {\mathbf w}_j
- v_{\parallel,j} {{\mathbf e}_{\rm v,nucl}}.
\end{eqnarray}
Taking into account Equations~\ref{eq:uvw_def_1} - \ref{eq:uvw_def_5}, Equation~\ref{eq:psi_def} reads
\begin{eqnarray}
 \!\!\! {\mathbf r} 
&=&  {\mathbf r}_{0,j} + \Psij \Delta{\mathbf v}\\ 
\nonumber &=&  {\mathbf r}_{0,j} 
+ \ve \, ( 
\sin \! \theta \sin \! \phi 
\underbrace{\Psij {\mathbf u}_j}_{\lambda \, {\mathbf e}_z}\!
+ \sin \! \theta \cos \! \phi \!\!\!
\underbrace{\Psij {\mathbf v}_j}_{\nu \, {\mathbf e}_y + \mu \,{\mathbf e}_z}\!
+ \cos \! \theta 
\underbrace{\Psij {\mathbf w}_j}_{{\mathbf c}})
- v_{\parallel,j} \underbrace{\Psij {{\mathbf e}_{\rm v,nucl}}}_{{\mathbf \sigma}}
\end{eqnarray}
or component-wise
\begin{eqnarray}
\label{eq:x}
x &=& x_{0,j} + \ve \, c_1 \cos \theta - v_{\parallel,j} \, \sigma_1 \\
\label{eq:y}
y &=& \underbrace{y_{0,j}  + \ve \, c_2 \cos \theta 
- v_{\parallel,j} \, \sigma_2}_{\tilde{y}_j (\theta)}
+ \underbrace{\ve \, \nu \sin \theta}_{\Delta y_{\rm max} (\theta)} 
\cos \phi.
\end{eqnarray}
These equations relate the direction of emission of a particle to its position
in the image. They can be inverted to give $\theta (x)$ and $\phi (x,y)$:
\begin{equation}
\label{eq:theta}
\cos \theta = \frac{v_{\parallel,j} \, \sigma_1 + (x \!-\! x_{0,j})}{\ve \, c_1}
\end{equation}
\begin{equation}
\label{eq:phi}
\cos \phi = \frac{y - \tilde{y}_j (\theta(x))}{\Delta y_{\rm max} (\theta(x))}.
\end{equation}
For fixed $\theta$, the full ring-segment in a pixel-column covers all values
of $\phi \in [-\pi,\pi]$. The borders of the projected shell are characterised by
$\phi=\pi$ and $\phi=0$, respectively. Hence, the two parts of a shell (front
and back) in the given pixel are described by the two angles $\phi(\theta)$ and
$-\phi(\theta)$. The sign of $\sin \phi$ does not change within a given side
of the shell.

\subsection{Parameters of the Map}
\label{subsec:math_method_params}

To exploit Equations~\ref{eq:theta} and \ref{eq:phi}, the parameters $\nu$,
$c_1$, $c_2$, $\sigma_1$, and $\sigma_2$ need to be evaluated.
First, the inverse matrix $\Psij^{-1}$ is calculated. The
column vectors in $\Psij^{-1}$ are denoted as
${\mathbf \Psi}^{-1}_{j (i)}$ with $i$ = 1 $\ldots$ 3 such that
\begin{equation}
\Psij^{-1} = 
({\mathbf \Psi}^{-1}_{j (1)}, {\mathbf \Psi}^{-1}_{j (2)}, {\mathbf \Psi}^{-1}_{j (3)})=
\left(\begin{array}{ccc}
\Psi^{-1}_{j 1,(1)} & \Psi^{-1}_{j 1,(2)} & \Psi^{-1}_{j 1,(3)}\\
\Psi^{-1}_{j 2,(1)} & \Psi^{-1}_{j 2,(2)} & \Psi^{-1}_{j 2,(3)}\\
\Psi^{-1}_{j 3,(1)} & \Psi^{-1}_{j 3,(2)} & \Psi^{-1}_{j 3,(3)}
\end{array}\right).
\end{equation}
Equation~\ref{eq:uvw_def_1} yields that  ${\mathbf u}_j$ = $\lambda
{\mathbf \Psi}^{-1}_{j (3)}$. With $|{\mathbf u}_j|$ = $1$ 
(Equation~\ref{eq:uvw_def_3}) follows
\begin{equation}
\lambda = \frac{1}{\sqrt{
(\Psi^{-1}_{j 1,(3)})^2 + (\Psi^{-1}_{j 2,(3)})^2 + (\Psi^{-1}_{j 3,(3)})^2}}.
\end{equation}
Equation~\ref{eq:uvw_def_2} gives 
\begin{equation}
\label{eq:v}
{\mathbf v}_j = \nu\, {\mathbf \Psi}^{-1}_{j (2)} + \mu\, {\mathbf \Psi}^{-1}_{j (3)}.
\end{equation}
Multiplication with ${\mathbf u}_j$ results in
\begin{equation}
0 = {\mathbf u}_j \, {\mathbf v}_j = 
\lambda \, \nu \, {\mathbf \Psi}^{-1}_{j (2)} \, {\mathbf \Psi}^{-1}_{j (3)} 
+ \lambda \, \mu \, \underbrace{({\mathbf \Psi}^{-1}_{j (3)})^2}_{1/\lambda^2},
\end{equation}
which can be solved for $\mu/\nu \, (\lambda, \Psij^{-1})$. Re-insertion into
  Equation~\ref{eq:v} gives the vector
  ${\mathbf v}_j/\nu$, from which $\nu$ is obtained due to the normalisation
  of ${\mathbf v}_j$. Then it is straightforward to calculate 
${\mathbf w}_j$ = ${\mathbf u}_j \times {\mathbf v}_j$, 
${\mathbf c}$ = $\Psij {\mathbf w}_j$, and
  ${\mathbf \sigma}$ = $\Psij \,{{\mathbf e}_{\rm v,nucl}}$. 

%% file: math_solution-general.tex
\subsection{General Solution}
\label{subsec:math_sol_gen}

The aim of this section is to calculate the number density of particles in a
given pixel confined by the coordinates ($x_1$, $y_1$) and ($x_2,y_2$). 
The distribution of particles released to a unit solid angle is given by
\begin{equation}
\frac{\df N}{\df \Omega} (\phi,\theta) = 
\frac{\df N(\phi,\theta)}{\df \phi \,\df \!\cos\theta},
\end{equation}
where $\theta (x)$ and $\phi (x,y)$ are the comet-centred angles defined in
Equations~\ref{eq:theta} and \ref{eq:phi}.
The amount of dust $N$ in this pixel is obtained by
\begin{eqnarray}
\nonumber
N &=&  \int\limits_{x_1}^{x_2} \int\limits_{y_1}^{y_2}
\frac{\df N}{\df x\,\df y} \df y\,\df x \\\nonumber
&=& \left| \, \int\limits_{\cos \theta_1(x_1)}^{\cos \theta_2(x_2)} 
\!\!\!\!\!\!\! \df \!\cos\theta
\int\limits_{\phi_1(y_1,\cos\theta)}^{\phi_2(y_2,\cos\theta)} 
\!\!\!\!\!\!\!\df \phi\
\frac{\df N}{\df \Omega}(\phi,\cos\theta) \right|
+ \left| \, \int\limits_{\cos \theta_1(x_1)}^{\cos \theta_2(x_2)} 
\!\!\!\!\!\!\! \df \!\cos\theta \!\!\!\!\!\!\!\! 
\int\limits_{-\phi_2(y_1,\cos\theta)}^{-\phi_1(y_2,\cos\theta)} 
\!\!\!\!\!\!\!\df \phi\
\frac{\df N}{\df \Omega}(\phi,\cos\theta)  \! \right|\!\\[3mm]
&=&  \left| N_1 \right| + \left| N_2 \right| .
\label{eq:N_ful}
\end{eqnarray}
The first integral represents the number density in the side of
the shell with $0 < \phi < \pi$ while the second integral represents the side
with negative $\phi$. Therefore, $N_1$ and $N_2$ cannot compensate each
other.

If a horizontal boundary of the pixel is inside the projected shell,
the corresponding limit of the inner integrals in Equation~\ref{eq:N_ful} is
given by 
Equation~\ref{eq:phi}. This is in the following referred to as a ``normal''
integration limit.
If the edge of the projected shell is inside the pixel, the limit $\pm
\phi_i\,(y_j, \cos \theta)$ of the inner integral must be set to $\phi$ = $0$ or
\mbox{$\phi$ = $\pi$} as appropriate (henceforth called an ``anomalous'' integration
limit). 
Introducing \mbox{$z$ = $\cos\theta$}, the inner
integration limits in Equation~\ref{eq:N_ful} are described by 
\begin{equation}
\phi(y,z) = \left\{
\begin{array}{ll}
 \phi_{\rm nm} (y,z) \hspace{2cm}& |\cos(\phi_{\rm nm})| < 1  \\ &\\
\phi_{\rm an} & \mathrm{else}
\end{array}\right. 
\end{equation} 
with $\phi_{\rm nm}(y,z)$ given by Equation~\ref{eq:phi}, and $\phi_{\rm an} \in \{0, \pi\}$.
The careful evaluation
of contributions from the edges of the projected shell is critical, because a
significant fraction of the dust can be concentrated in these pixels
due to the shallow angle between the line of sight and the surface of the
shell.  

\begin{figure}[t]
\center
\includegraphics[clip,width=\textwidth]{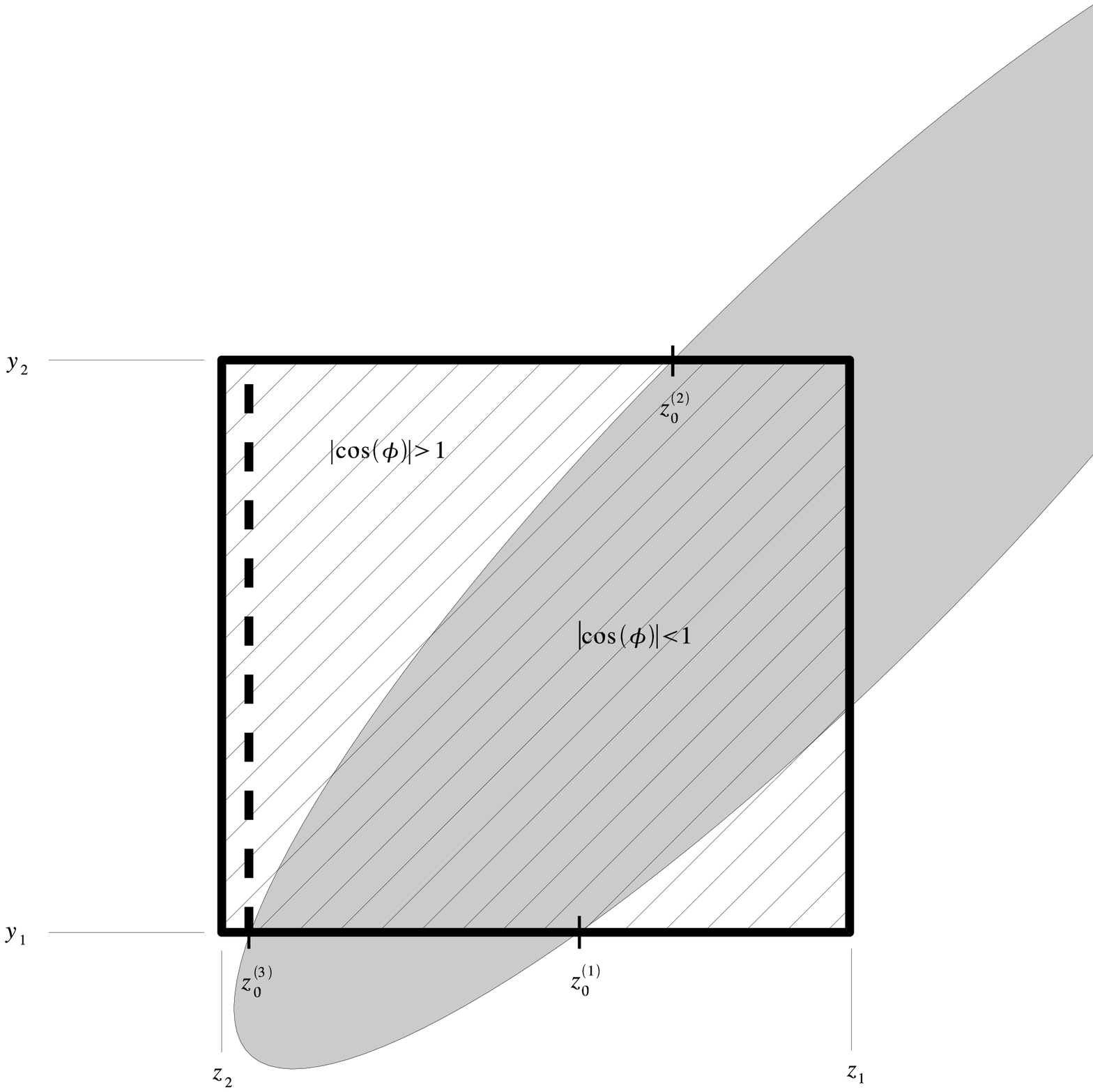}
\caption{Example for a pixel (hatched area) containing part of the edges of a
  shell (area covered by shell is grey). The $z_0^{(i)}$ are defined by
   \mbox{$z_0^{(i)}$ = $z(|\cos \phi| = 1, y_j)$}. The integration limits
  for this pixel are as follows: on the upper pixel boundary,
   $\phi$ = $\phi_{\rm nm}(y_2,z)$ for $z \in [z_1, z_0^{(2)}]$, 
   while $\phi$ = $\phi_{\rm an}(y_2)$ for the rest. 
   On the lower pixel boundary, $\phi$ = $\phi_{\rm  nm}(y_1,z)$ 
   for $z \in [z_0^{(1)}, z_0^{(3)}]$, and 
   $\phi$ = $\phi_{\rm an}(y_1)$ for $z \in [z_1, z_0^{(1)}]$. 
   For $z \in [z_0^{(3)}, z_2]$, neither boundary contributes. Hence this
   section does  not have to be considered at all. This effectively
   corresponds to a shortening of the 
  pixel, indicated by the dashed vertical line.}
\label{fig:boundary_example}
\end{figure}
\begin{figure}[h]
\center
\includegraphics[clip,angle=270,width=\textwidth]{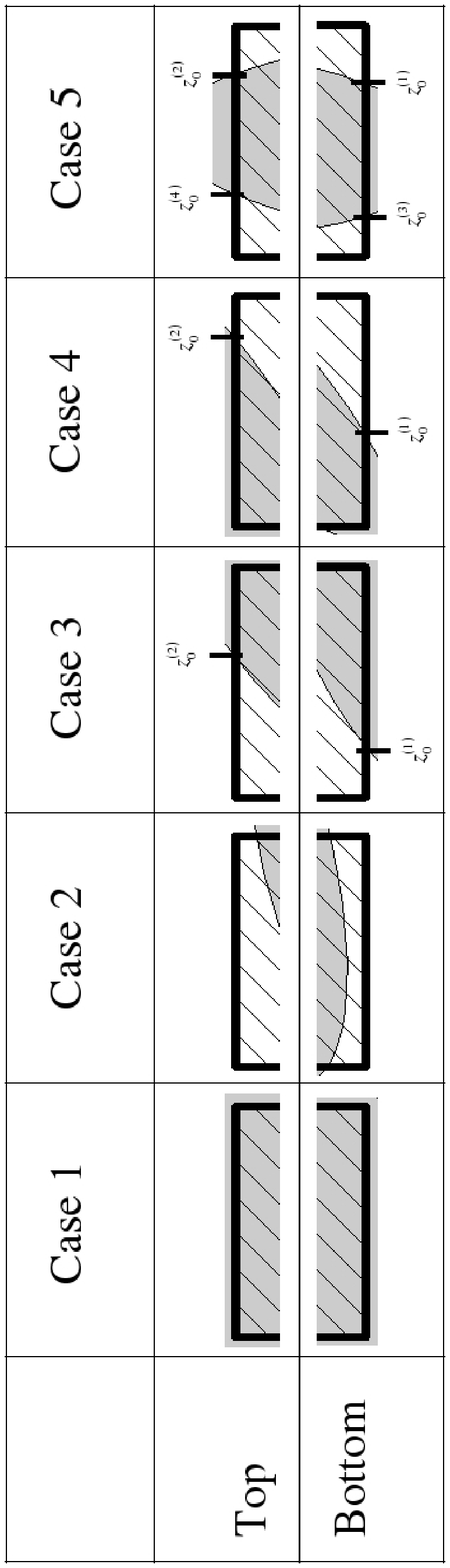}
\caption{Different possibilities of pixel boundaries intersecting the edge of
a shell. The
area covered by the shell is grey. The
upper and lower rows show situations for the upper and lower
 pixel boundaries, respectively. Different ``cases'' occur independently for
the upper and lower pixel boundary.
{\bf 1}:
  Both corners are covered by the shell. {\bf 2}: Both corners are outside the
  shell. If this scenario is given on both pixel boundaries, the pixel is either empty ($\phi_1$ =
  $\phi_2$) or the shell is thinner than the height of the pixel 
  ($\phi_{1/2}$ = $\pi$ and $\phi_{2/1}$ = $0$). 
  This situation occurs for shells that are
  collapsed to a neckline.
  {\bf 3}: The right corner is inside, the left corner outside the shell. 
  {\bf 4}: The inverse of case 3.
  {\bf 5}: Both corners are outside the shell, but there are two shell
  intersections in between. The possible sixth case (inverse of 5) cannot
  occur because shells are convex. 
In all cases, if there is only one intersection, the
  intersection point is labelled $z_0^{(1)}$ on the lower pixel boundary and
  $z_0^{(2)}$ on the upper one. 
If there is a second intersection, the
  leftmost intersection point is labelled $z_0^{(3)}$ on the lower
  and $z_0^{(4)}$ on the upper boundary.}
\label{fig:boundary_cases}
\end{figure}

As an example, the integral $N_1$ is given for the situation depicted in
Figure~\ref{fig:boundary_example}.
Introducing $dN/d\Omega$ = $f(\phi,z)$ and its antiderivative $F(\phi,z)$, the
integral $N_1$ is  
\begin{eqnarray}
\nonumber
N_1 
&=& \int\limits_{z_1}^{z_2} \df z \int\limits_{\phi_1(y_1,z)}^{\phi_2(y_2,z)}
\df \phi \, f(\phi,z)
\\\nonumber
&=& \int\limits_{z_1}^{z_2} \df z \left[ F(\phi_2(y_2,z),z) - F(\phi_1(y_1,z),z)
  \right]
\\\nonumber
&=& \int\limits_{z_1}^{z_0^{(2)}} \df z F(\phi_{\rm nm}(y_2,z),z) +
\int\limits_{z_0^{(2)}}^{z_0^{(3)}} \df z F(\phi_{\rm an}(y_2,z_2),z)
\\
&& - \int\limits_{z_1}^{z_0^{(1)}} \df z F(\phi_{\rm an}(y_1,z_1),z) -
\int\limits_{z_0^{(1)}}^{z_0^{(3)}} \df z F(\phi_{\rm nm}(y_1,z),z).
\label{eq:bound_cases_example}
\end{eqnarray}
For each horizontal pixel boundary, five scenarios of intersection with the
edge of the shell 
can be distinguished. These are depicted in Figure~\ref{fig:boundary_cases}.

%% file: math_solution-hom-emission.tex
\subsection{Isotropic Dust Emission} 
\label{subsec:math_sol_hom}
Isotropic emission is described by $f = \df N / \df \Omega = Q/4\pi$. In this instance, the
integral in Equation~\ref{eq:N_ful} can be solved analytically. 
For symmetry
reasons, the contributions from the front and back sides of the shell are
identical, and Equation~\ref{eq:N_ful} simplifies to 
\mbox{$N^{\rm (iso)}$ = $2 \,N_1^{\rm (iso)}$} with
\begin{equation}
N_1^{\rm (iso)} = 
\int\limits_{z_1}^{z_2} 
\!\!\df \!z \!\!\!\!
\int\limits_{\phi (y_1,z)}^{\phi (y_2,z)} 
\!\!\!\!\!\!\df \phi\
\frac{\df N}{\df \Omega}(\phi,z) 
= 
\frac{Q}{4\pi} \int\limits_{z_1}^{z_2} 
\!\!\df \!z 
\left[ \phi(y_2,z) - \phi(y_1,z) \right]
= N_{1a}^{\rm (iso)} - N_{1b}^{\rm (iso)}, 
\label{eq:N1h}
\end{equation}
where $z_1$ = $z(x_1)$, $z_2$ = $z(x_2)$, and ``a''
and ``b'' refer to the upper and lower pixel boundary 
($y$ = $y_2$ and $y$ = $y_1$), respectively. 

\vspace{\baselineskip}
\noindent{\bf Normal integration limits.} 
If the part of the upper pixel boundary that is inside the shell is limited by
$z_{\alpha}$ and $z_{\beta}$ (in the example given in
Equation~\ref{eq:bound_cases_example} 
$z_{\alpha}$ = $z_1$ and $z_{\beta}$ = $z_0^{(2)}$), the corresponding normal part of the integral $N_{1a}^{\rm (iso)}$ is given by 
\begin{eqnarray}\nonumber
N_{1a}^{\rm (iso)} &=& 
\frac{Q}{4\pi} \int\limits_{z_{\alpha}}^{z_{\beta}} 
\df z \,\, \phi_{\rm nm} (y_2,z)\\[3mm]\nonumber
&=& 
\frac{Q}{4\pi} \int\limits_{z_{\alpha}}^{z_{\beta}} 
\df z \,\, 
\arccos \left( \frac{y_2 - \tilde{y}_j (\theta(x))}{\Delta y_{\rm max}
  (\theta(x))} \right)\\[3mm]\nonumber
&=& 
\frac{Q}{4\pi} \int\limits_{z_{\alpha}}^{z_{\beta}} 
\df z \,\, 
\arccos \left( \frac{A_2 + Bz}{\sqrt{1-z^2}} \right)
\\[3mm]\nonumber
&=& \frac{Q}{4\pi}
\left[
z \arccos\left( 
\frac{A_2+Bz}{\sqrt{1-z^2}} \right)\right.\\[3mm]\nonumber
&& \left. - \frac{1}{2}  
\arctan \left( \frac{A_2^2+A_2Bz-1 - p_{2,2}}{(A_2-B)\sqrt{p_{1,2}}}
\right) \right.\\[3mm]\nonumber
&& \left. + \frac{1}{2} 
\arctan \left( \frac{A_2^2+A_2Bz-1 + p_{2,2}}{(A_2+B)\sqrt{p_{1,2}}}
\right) \right.\\[3mm]
&& \left. - \frac{A_2}{\sqrt{1+B^2}}\
\arctan \left( \frac{p_{2,2}}{\sqrt{1+B^2}\sqrt{p_{1,2}}} \right) 
\right]_{z_{\alpha}}^{z_{\beta}} ,
\label{eq:N_1a_iso}
\end{eqnarray}
with
\begin{eqnarray}
A_2 &=& \frac{y_2 - y_{j, 0} + v_{\parallel, j}\sigma_2}{\ve \nu}
\label{eq:A-def}\\
B &=& -\frac{c_2}{\nu}
\label{eq:B-def}\\
p_{1,2} &=& 1 - z^2 - (A_2+Bz)^2
\label{eq:p1-def}\\
p_{2,2} &=& z + B^2 z + A_2B.\label{eq:p5-def}
\end{eqnarray}
The second index ``2'' indicates that $A_2$, $p_{1,2}$, and $p_{2,2}$ depend
on the upper pixel boundary ($y$ = $y_2$). 

If the edge of the shell intersects the pixel boundary, one or both of 
$z_{\alpha}$ and $z_{\beta}$ in Equation~\ref{eq:N_1a_iso} are given by an
intersection point rather than by a pixel corner.
As discussed at the end of
Section~\ref{subsec:math_method_projection}, for the value of $z$
in question follows $\cos \phi_{1/2}$ = $\pm$ 1, and some
simplifications can be introduced into Equation~\ref{eq:N_1a_iso}.
To avoid numerical errors at the intersection
points, it is advisable to
introduce these simplifications explicitely into the computer code.
\begin{eqnarray}
\label{eq:simp_cosphi}
\frac{A_2+Bz_{\alpha/\beta}}{\sqrt{1-z_{\alpha/\beta}^2}} &=& \pm 1 = k_{\rm cos},\\
p_{1,2} &=& 0,\\
p_{2,2} &=& z_{\alpha/\beta} + k_{\rm cos} B \sin\theta,\\
\label{eq:simp_arctan}
\arctan(\frac{G}{\sqrt{p_{1,2}}}) &=& \frac{\pi}{2} \, {\rm sign}(G).
\end{eqnarray}
The expression $G/\sqrt{p_{1,2}}$ refers to any of the three arguments of the $\arctan$ in the
last three lines of Equation~\ref{eq:N_1a_iso}.
The simplified expression for the concerned term in $N_{1a, \alpha/\beta}^{\rm (iso)}$ then reads
\begin{eqnarray}\nonumber
N_{1a, \alpha/\beta}^{\rm (iso)} 
&=& \frac{Q}{4\pi} \left[ k z_{\alpha/\beta} 
\right.\\[3mm]\nonumber
&& \left. - \, \frac{\pi}{4} \,   
{\rm sign} \left( \frac{A_2^2+A_2Bz_{\alpha/\beta}-1 - p_{2}}{(A_2-B)}
\right) \right.\\[3mm]\nonumber
&& \left. + \, \frac{\pi}{4} \, 
{\rm sign} \left( \frac{A_2^2+A_2Bz_{\alpha/\beta}-1 + p_{2}}{(A_2+B)}
\right) \right.\\[3mm]
&& \left. - \frac{A_2}{\sqrt{1+B^2}} \, \frac{\pi}{2} \,
{\rm sign} \left( p_{2} \right) 
\right].
\end{eqnarray}
The solution of $N_{1b}^{\rm (iso)}$ is analogous to $N_{1a}^{\rm (iso)}$,
replacing $y_2$ by $y_1$, and in
general with different values for $z_{\alpha}$ and $z_{\beta}$. In the example
given in
Equation~\ref{eq:bound_cases_example} the lower boundary would be
characterised by $z_{\alpha} =
z_0^{(1)}$ and $z_{\beta} = z_0^{(3)}$.

\vspace{\baselineskip}
\noindent{\bf Anomalous integration limits.}
For those sections of a pixel boundary that are outside the shell (limited by
$z_{\gamma}$ and $z_{\delta}$), 
the inner integration limit in Equation~\ref{eq:N1h} simplifies to
\mbox{$\phi (y_i,z)$ = $\phi_{\rm an}=k_i$}, 
where the index ``$i$'' refers to the
upper and lower pixel boundary ($y_2$ or $y_1$), and the value of $k_i$ 
is either 0 or $\pi$. 
In the example given in Equation~\ref{eq:bound_cases_example}, 
$z_{\gamma}$ = $z_0^{(2)}$ and $z_{\delta}$ = $z_0^{(3)}$ for the upper, and
$z_{\gamma}$ = $z_1$ and $z_{\delta}$ = $z_0^{(1)}$ for the lower pixel
boundary. 
The anomalous part of the integral $N_{1a/b}^{\rm (iso)}$ in
Equation~\ref{eq:N1h} reads then
\begin{eqnarray}\nonumber
N_{1a}^{\rm (iso)} &=& 
\frac{Q}{4\pi} \int\limits_{z_{\gamma}}^{z_{\delta}} 
\df z \,\, \phi_{\rm an} (y_2,z)\\[3mm]\nonumber
&=& 
\frac{Q}{4\pi}k_2
\left[ z_{\delta} - z_{\gamma}  \right].
\end{eqnarray}
As for normal boundaries, the solution for the lower pixel boundary, $N_{1b}^{\rm (iso)}$, is analogous to $N_{1a}^{\rm (iso)}$,
replacing $y_2$ by $y_1$, and having different values for $z_{\gamma}$ and
$z_{\delta}$. 